\documentclass[12pt,preprint]{aastex}
\usepackage{multirow,amsthm,amsmath,amssymb,mathrsfs}
\usepackage{graphicx,verbatim,color,ulem,gensymb,arydshln}

\def\aap{Astron.\ Astrophys.\ }

\def\apjl{Astrophys.\ J.\ Lett.\ }
\def\apjs{Astrophys.\ J.\ Supp.\ }

\newcommand*{\si}[1]{{\rm {#1}}}

\graphicspath{{./}{figs/}}

\begin{document}

\title{Galactic diffuse emission from radio to ultra-high-energy $\gamma$ rays in light of up-to-date cosmic ray measurements}

\author{Xin-Yu He$^{1,2,3}$, Pei-Pei Zhang$^{1,2,3}$, Qiang Yuan$^{1,2}$, Yi-Qing Guo$^{3,4}$}

\affil{$^1$Key Laboratory of Dark Matter and Space Astronomy, Purple Mountain Observatory, Chinese Academy of Sciences, Nanjing 210023, China; yuanq@pmo.ac.cn\\
$^2$School of Astronomy and Space Science, University of Science and Technology of China, Hefei 230026, China\\
$^3$Key Laboratory of Particle Astrophysics, Institute of High Energy Physics, Chinese Academy of Sciences, Beijing 100049, China; guoyq@ihep.ac.cn\\
$^4$University of Chinese Academy of Sciences, Beijing 100049, China
}

\begin{abstract}

Cosmic rays travel throughout the Galaxy, leaving traces from radio to ultra-high-energy $\gamma$ rays due to interactions with the interstellar gas, radiation field and magnetic field. Therefore, it is necessary to utilize multi-wavelength investigations on the Galactic diffuse emission to shed light on the physics of CR production and propagation. In this work, we present a spatially dependent propagation scenario, taking account of a local source contribution, while making allowances for an additional CR component freshly accelerated near their sources. In this picture, after reproducing the particle measurements at the Solar system, we calculated the intensity and compared the spectral energy distribution to observations from Fermi-LAT and LHAASO-KM2A in the $\gamma$-ray band, and from WMAP and Planck among other radio surveys at lower energies. Multi-band data considered in conjunction, the former comparison exhibits sufficiently good consistency in favor of our model, while the latter calls for improvement in data subtraction and processing. From this standpoint, there remains potential for advanced observations at energies from milli-$\si{eV}$s to $\si{MeV}$s towards the Galactic plane, in order to evaluate our model further and more comprehensively in the future.

\end{abstract}

\keywords{Cosmic rays --- Galactic diffuse emission ---Astrophysics - High Energy Astrophysical Phenomena}

\label{firstpage}


\section{Introduction}

Interactive and ubiquitous, cosmic rays (CRs) play an impactful role in varieties of celestial events. Nonetheless, the origin of CRs remains a century-long mystery. CRs below the knee (around a few $\si{PeV}$s) are expected to be of Galactic origin. Away from their energetic accelerators, particles propagate in the magnetic halo of the Milky Way and interact with the interstellar gas, radiation field (ISRF) and magnetic field (IMF), generating secondary particles and photons. Unlike the particle measurements, which lose most of their initial directional information, the secondary photon emission preserves the spatial distribution of the progenitor CRs and thus turns out to be a unique and irreplaceable probe of CR propagation.

Wide-band diffuse emission from radio frequencies to ultra-high-energy (UHE) $\gamma$ rays is yielded via interactions between CRs and the interstellar gas, the ISRF, and the IMF. In general, for a typical random magnetic field of a few $\si{\mu G}$s, the synchrotron radiation of CR electrons and positrons (CREs) results in radio emission from $\si{MHz}$ to $\si{THz}$. The bremsstrahlung of CREs in the ISM generates high-energy emission from X rays to $\gamma$ rays. The inverse Compton scattering (ICS) between CREs and the ISRF, together with the inelastic hadronic interactions between CR nuclei and the ISM give diffuse $\gamma$ rays in a wide energy range. The multi-wavelength diffuse emission observations can therefore be used to constrain the source distribution and propagation of Galactic CRs.

With forefront space-borne and ground-based instruments, $\gamma$-ray observations have advanced into higher energy domains, enabling the possibility to study the GDE in multi-wavelength approaches. Radio \citep{1981Natur.289..470H, 2015MNRAS.451.4311R} and microwave \citep{2016A&A...594A..10P, 2016A&A...594A..25P} surveys of early years, in conjunction with recent measurements at above hundreds of $\si{MeV}$s \citep{2012ApJ...750....3A} and at even higher energies of multi-$\si{TeV}$s \citep{2015ICRC...34..838L, 2015ICRC...34..787S} can be comprehensively investigated to renovate and reconstruct the current theoretical framework of CRs. While some previous studies have taken data-driven, phenomenological methodologies, others have proposed variant refined models beyond our traditional understanding of CRs, postulating exotic origins \citep{2007PhRvD..75h3006B} and/or novel interaction mechanisms \citep{2015JCAP...03..038C}, but only a few of them show the capability to interpret the observed GDE spectrum across different wavelengths and other unforeseen anomalous phenomena simultaneously \citep{2010ApJ...722L..58S, 2016PhRvD..94f3504C, 2016JPhCS.718e2027O}. Straightforward comparison of high-level models each derived from within a certain energy range corresponding to a certain series of astronomical entities and processes could sometimes be misleading, due to the interdependence among the involved physical quantities and uncertainties introduced by different assumptions. Thus, it might be a safer and more refreshing perspective that we aim at fitting all available data at different energy ranges inside a unified configuration when trying to construct or assess our models, which partially motivates this work. Meanwhile, multi-wavelength observations have been used in many recent studies on both point and extended sources, which rely heavily on the accurate modeling of the GDE.

Featuring a spatially dependent propagation (SDP) scenario with extra CR origins beyond the standard paradigm, our model, which has been developed based on principally the up-to-date measurements of secondary-to-primary ratios \citep{2023JCAP...02..007Z}, could be finer tested on this wise. In recent years, it has been argued extensively that the excesses in the observed CR spectra could be naturally explained by a rather young supernova remnant (SNR) located near us \citep{2016ApJ...819...54G, 2018ApJ...869..176L},  whose contribution is also regarded in this work. The suggested CR confinement around the accelerators also influences the GDE intensity across different frequencies and will be depicted more specifically hereafter.

The rest of this paper is organized as follows. Section~\ref{mdl} describes the model setting and obtains the model parameters. Section~\ref{gde} presents and discusses the results of wide-band diffuse emission. Finally, Section~\ref{cnc} concludes this work.

\section{Model}
\label{mdl}
During the active phases of varieties of astrophysical objects, particles are accelerated up to very high energies (VHE) and then go through a long voyage across the Galaxy before some of them enter the solar system. The whole process can be described from three major aspects: the injection from sources, the propagation in the Milky Way, and the production of secondaries.

\subsection{Source injection}
\label{src} 
In this work, the SNRs are considered as the dominant sources of CRs and a continuous distribution \citep{1996A&AS..120C.437C} is adopted.
The spatial distribution can be parameterized as 
\begin{equation}
    f(r,z)=\left(\frac{r}{r_\odot}\right)^a \exp\left[-b\frac{(r-r_\odot)}{r_\odot}\right]\exp\left(-\frac{|z|}{z_s}\right),
\end{equation}
where $r_\odot=8.5\ \si{kpc}$ is the distance of the Solar system to the Galactic center. Other parameters are adopted as $a=1.69$, $b=3.33$, and $z_s=0.2\ \si{kpc}$ \citep{1996A&AS..120C.437C}. Besides this constituent defined as the background component, the contribution from individual nearby sources is also considerable, as evidenced by the curiosities in the energy spectra \citep{2017ApJ...839....5Y, 2018JETPL.108....5A, 2019PhRvL.122r1102A, 2019SciA....5.3793A} and large-scale anisotropies \citep{2015ApJ...809...90B, 2016ApJ...826..220A, 2017ApJ...836..153A, 2019ApJ...871...96A}. To reproduce these anomalies, a local source is incorporated to our model. 
Though many astrophysical objects exhibit the ability to accelerate particles up to VHE, the SNRs are generally considered as the dominant ones. The injection spectra are assumed to follow a broken power-law energy distribution:
\begin{equation}
    Q_{\rm pri}(\mathcal{R})=Q_0 \times
    \begin{cases}
        (\frac{\mathcal{R}}{\mathcal{R}_{br}})^{-\nu_1} &, \ \mathcal{R}\leq\mathcal{R}_{\rm br}\\
        (\frac{\mathcal{R}}{\mathcal{R}_{br}})^{-\nu_2}\exp[-\frac{\mathcal{R}}{\mathcal{R}_c}] &,\ \mathcal{R}>\mathcal{R}_{\rm br}\\
    \end{cases},
\end{equation}
where $Q_0$ is the normalization factor, $\nu_1$ ($\nu_2$) is the spectral index at rigidities lower (higher) than the break rigidity $\mathcal{R}_{\rm br}$, and $\mathcal{R}_c$ is the cut-off rigidity. The detailed information of the injection spectra is listed in Table~\ref{tab:bg_inj} and Table~\ref{tab:loc_inj} for the background and the local source respectively. 
\begin{table}[htb]
	\centering
	\caption{Galactic background injection parameters.}
	\label{tab:bg_inj}
	\begin{tabular}{ccccccc}\\
	    \hline\hline
		$\Phi[\si{MV}]$ & Species & $Q_0\ [\si{m^{-2}sr^{-1}s^{-1}GeV^{-1}}]$ & $\nu_1$ & $\nu_2$ & ${\mathcal R}_{\rm br}\ [\si{GV}]$ & ${\mathcal R}_c$\ [\si{TV}]\\ \hline
		550 & Electron & {$2.35 \times 10^{-1}$} & 1.90 & 2.60 & 4.25 & 10\\ 
		    & Proton   & {$4.40 \times 10^{-2}$} & 1.40 & 2.38 & 5.50 & 1600\\
		\hdashline
		400 & Electron & {$2.35 \times 10^{-1}$} & 1.99 & 2.60 & 12.25 & 10\\ 
		    & Proton   & {$4.44 \times 10^{-2}$} & 1.60 & 2.38 & 8.77 & 1600\\
		\hdashline
		700 & Electron & {$2.35 \times 10^{-1}$} & 1.50 & 2.60 & 0.50 & 10\\ 
		    & Proton   & {$5.25 \times 10^{-2}$} & 1.15 & 2.38 & 4.00 & 1600\\
		\hline\hline
	\end{tabular}
\end{table}

\begin{table}[htb]
	\centering
	\caption{Local SNR injection parameters.}
	\label{tab:loc_inj}
	\begin{tabular}{ccccc}\\
		\hline\hline
		Species & $q_0[\si{m^{-2}sr^{-1}s^{-1}GeV^{-1}}]$ & $\gamma$ & $\mathcal{R}_c[\si{TV}]$\\
		\hline
		Electron & $1.2 \times 10^{50}$ & 2.10 & 10\\
		Proton   & $5.0 \times 10^{52}$ & 2.08 & 30\\
		Carbon   & $4.0 \times 10^{50}$ & 2.08 & 15\\
		\hline\hline
	\end{tabular}
\end{table}

\subsection{CR propagation}

As mentioned above, we adopt an SDP model to describe the propagation of CRs in the Milky Way.
The SDP model is primarily motivated\footnote{An original motivation of the SDP model is to explain the hardenings of CR spectra \citep[e.g.][]{2012ApJ...752L..13T}. However, this scenario alone may not be easy to account for new observations of softening features of CR spectra around 10 TeV \citep{2019SciA....5.3793A}.} by $\gamma$-ray observations of pulsar halos, which suggest very slow diffusion for regions surrounding those pulsars \citep[e.g.][]{2017Sci...358..911A, 2021PhRvL.126x1103A} compared with the average diffusion coefficient inferred from CR secondary-to-primary ratios. It is a natural expectation that the slow diffusion regions in the Galactic plane are abundant due to the population of such middle-aged pulsars. 
Therefore, a two-zone diffusion (slow disk plus fast halo) scenario is reasonable to describe the propagation of CRs. It was also shown that this two-zone diffusion model can suppress the amplitudes of the dipole anisotropies, and the spatial variations of CR intensities and spectra from Fermi observations \citep{2018PhRvD..97f3008G}. 

Following the previous work \citep{2018PhRvD..97f3008G}, we assume an anti-correlation between the diffusion coefficient with the source distribution. The diffusive halo is divided into two regions: the slow diffusion inner halo (IH) and the fast diffusion outer halo (OH). At the outer halo border ($r=r_h$, $z=\pm z_h$), the free-escape condition, $\psi(r_h,z,p)=\psi(r,\pm z_h,p)=0$, is imposed. With $z_h$, $\xi z_h$, and $(1-\xi)z_h$ being the half-thickness of the entire halo, the IH, and the OH respectively, the diffusion coefficient is given by
\begin{equation}
    D(r,z,\mathcal{R})=D_0F(r,z)\beta^\eta\left(\frac{\mathcal{R}}{\mathcal{R}_0}\right)^{\delta_0F(r,z)},
\end{equation}
where $\beta$ is the velocity of the CR particle in unit of light speed $c$, and
\begin{equation}
    F(r,z)=
    \begin{cases}
        g(r,z)+[1-g(r,z)]\left(\frac{z}{\xi L}\right)^n, & |z|\leq\xi z_h\\
        1, & |z|>\xi z_h
    \end{cases},
\end{equation}
where $g(r,z)=\frac{N_m}{1+f(r,z)}$. The propagation parameters are summarized in Table~\ref{tab:prop_param}. For a more complete and detailed description of the SDP model, we refer the readers to \citet{2023JCAP...02..007Z}.

\begin{table}[htb]
	\centering
	\caption{Propagation parameters of the SDP model.}
	\label{tab:prop_param}
	\begin{tabular}{ccccccccc}
		\hline\hline
		$D_0\ [\si{cm^{-2}s^{-1}}]$ & $\delta_0$ & $N_m$ & $\xi$ & $n$ & $\eta$ & $\mathcal{R}_0$ & $z_h\ [\si{kpc}]$ & $v_A\ [\si{kms^{-1}}]$\\
		\hline
		$5.1 \times 10^{28}$ & 0.56 & 0.55 & 0.1 & 4.0 & 0.05 & 4.0 & 5.0 & 6.0\\
		\hline\hline
	\end{tabular}
\end{table}

As CRs enter the heliosphere, their energy spectra will be modified by the solar magnetic field. This so-called solar modulation effect is accounted for using the prevalent force field approximation \citep{1968ApJ...154.1011G}. In this work, without considering the charge-sign dependent modulation effect, a constant modulation potential of $\sim 550\pm 150$ MV is assumed, which, together with other parameters, suffices to fit the observed CR spectra properly. However, it should be noted that this simplified treatment results in uncertainties of the calculated diffuse emission, particularly for the low-frequency (radio) and low-energy ($\gamma$-ray) parts.

\subsection{Secondary production}
\label{prod_sec}

As CRs travel through the Milky Way, they leave substantial imprints of secondary nuclei, leptons (positrons and electrons), and photon emission of radio to $\gamma$ rays, which provide important probes to study their propagation. 
Relatively heavy secondary nuclei, such as lithium, beryllium and boron, are produced through fragmentation of the primaries chancing upon interstellar gas molecules, in which case each nucleon is considered to take after the energy of its parent. The production of these secondary particles can be described as follows:
\begin{equation}
    Q_{{\rm sec},j}=\sum_{i>j}(n_\mathrm{H}\sigma_{i+\mathrm{H}\rightarrow j}+n_{\mathrm{He}}\sigma_{i+\mathrm{He}\rightarrow j})v_i\psi_i,
\end{equation}
where $\psi_i$ is the density of the primaries, $v_i$ is the velocity of the parent particle, $n_{\mathrm{H},\mathrm{He}}$ is the number density of hydrogen/helium in the ISM, and $\sigma_{i+\mathrm{H}/\mathrm{He}\rightarrow j}$ denotes the cross section of the fragmentation process $i\rightarrow j$. 

Besides the fragmentation of heavy nuclei, inelastic collisions of light nuclei with the ISM will also produce secondary particles, such as antiprotons, electrons, positrons, and $\gamma$ rays. The source term is the convolution of the primary spectra and the relevant differential cross sections:
\begin{equation}
    W_{{\rm sec},j}(p_j)=\sum_{i=\mathrm{p},\mathrm{He}}\int dp_iv_i\left[n_\mathrm{H}\frac{d\sigma_{i+\mathrm{H}\rightarrow j}(p_i,p_j)}{dp_j}+n_{\mathrm{He}}\frac{d\sigma_{i+\mathrm{He}\rightarrow j}(p_i,p_j)}{dp_j}\right]\psi_i(p_i),
\end{equation}
where $\psi_i(p_i)$ is the solution of the propagation equation of the primaries in the Milky Way. The interactions near the acceleration sites are also considered, in which case, we assume that these particles have not experienced adequate propagation before they escape from the source regions, and $\psi_i=Q_{{\rm pri},i}\times\tau$, where $\tau$ is the effective confinement time of the particles around the source, which is estimated to be 0.2 Myr in our model through fitting the observed CR spectra. Approximately, the average grammage accumulated by all escaped primary particles from $t=0$ to $t=\tau$ is $\bar{X}=\rho c\tau \approx m_pn_{\rm H}c\tau=0.32\ \si{g~cm^{-2}}$, assuming a constant density $n_{\rm H}=1$~cm$^{-3}$ in the proximity of the sources for such injections.
The charged secondary particles experience the same propagation procedure as primary CRs in the Milky Way.

The $\gamma$-ray emissivity from $pp$ collisions can also be calculated with Eq. (6). For the production cross section, we use a recently developed interpolation routine based on Monte Carlo simulations, {\tt AAfrag}, which employs the updated QGSJET-II-04m model tuned with the LHC data on hadronic interactions \citep{2019CoPhC.24506846K}. For proton energies below 4 GeV and helium energies below 5 GeV, where {\tt AAfrag} does not cover, an old cross section model of \citet{1986A&A...157..223D} is used. 

The bremsstralung emissivity of CREs is 
\begin{equation}
    \frac{dN}{d\epsilon}=\alpha r_0^2 c \epsilon^{-1} \sum_s n_s \int_{max(\epsilon,E_{e,min})}^{E_{e,max}}dE_e \psi_e(E_e) E_e^{-2}\times [(2E_e^2-2E_e\epsilon+\epsilon^2)\phi_{1,s}-2E_e(E_e-\epsilon)\phi_{2,s}/3],
\end{equation}
where $\alpha$ is the fine structure constant, $r_0$ is the classical electron radius, $c$ is the speed of light, $n_s$ is the number density of gas of species $s$, $E_e$ is the energy of CREs, $\psi_e$ is the differential density of CREs, $\epsilon$ is the energy of emitted photons, $\phi_{1,s}$ and $\phi_{2,s}$ are shielding factors which can be found in \citet{1970RvMP...42..237B}.

The gas distributions used in this work are the ones embedded in GALPROP. They are HI distribution from \citet{1976ApJ...208..346G} renormalized to the results given by \citet{1990ARA&A..28..215D}, the molecular gas distribution from \citet{1988ApJ...324..248B} with a conversion factor from CO to H$_2$ of $\sim1.9\times10^{20}$ mols cm$^{-2}$/(K km s$^{-1}$) \citep{1996A&A...308L..21S}, and the ionized hydrogen distribution from \citet{1991Natur.354..121C}.
As for the calculation of the diffuse emission, the gas column densities for each line of sight are further renormalized to match the data from surveys of HI \citep{2005A&A...440..775K} and molecular gas \citep{2001ApJ...547..792D}.

In addition to the interactions depicted above, CREs also emit $\gamma$ rays via the ICS process in the ISRF. The ICS $\gamma$-ray emissivity is \citep{2000ApJ...528..357M}
\begin{equation}
    \frac{dN}{d\epsilon_2}=\frac{n_en_\gamma c}{4\pi m_e c^2}\int{d\epsilon_1}\int{d\gamma \epsilon_1^2 \gamma^2 f_\gamma(\epsilon_1) f_e(\gamma) \frac{dR(\gamma,\epsilon_1)}{d\epsilon_2}},
\end{equation}
where $\gamma$ is the Lorentz factor of parent electrons, $\epsilon_1$ is the energy of seed photons, $\epsilon_2$ is the energy of scattered photons, $(n_\gamma \epsilon_1^2 f_\gamma)$ and $(n_e\gamma^2f_e)$ are the differential spectra of seed photons and parent electrons, and
\begin{equation}
    \frac{dR}{d\epsilon_2}(\gamma,\epsilon_1)=\frac{2\pi r^2_e}{\epsilon_1\gamma^2}[2q^\prime \ln{q^\prime}+(1+2q^\prime)(1-q^\prime)+\frac{1}{2}\frac{(4\epsilon_1\gamma q^\prime)^2}{(1+4\epsilon_1\gamma q^\prime)}(1-q^\prime)]
\end{equation}
is the yield function of ICS photons for electron-photon interactions of fixed energies, for an isotropic distribution of photons, where $q^\prime=\epsilon_2/[4\epsilon_1\gamma^2(1-\epsilon_2/\gamma)]$ and $1/4\gamma^2<q^\prime\leq1$.

VHE $\gamma$ rays would experience attenuation when traveling through the ISRF due to the pair production effect \citep{2006A&A...449..641Z,2006ApJ...640L.155M}. We calculate the absorption fraction of the three dimensional emissivity of $\gamma$ rays, and then integrate along the line-of-sight to get the fluxes \citep{2023ApJ...957...43Z}.

The Milky Way is filled with randomly oriented magnetic field, and the electrons will lose energy and produce wide-band emission through the synchrotron process. After averaging over the pitch angle for an isotropic electron distribution, the synchrotron emissivity of a single electron integrated over all directions is \citep{1988ApJ...334L...5G}
\begin{equation}
    \epsilon(\nu)=4\sqrt{3}\pi r_e m_e c\nu_B x^2\{K_{4/3}(x)K_{1/3}(x)-3x[K_{4/3}(x)K_{4/3}(x)-K_{1/3}(x)K_{1/3}(x)]/5\},
\end{equation}
where $\nu$ is the radiation frequency, $\gamma$ is the electron Lorentz factor, $\nu_B = eB/(2m_e c)$,  $x=\nu/(3\gamma^2 \nu_B)$, and $K_{4/3}$, $K_{1/3}$ are modified Bessel functions.
The Galactic magnetic field strength is modeled as $B=B_0e^{-R/30\,{\rm kpc}-|z|/4\,{\rm kpc}}$. The observable synchrotron intensity is then obtained by integrating the emissivity along the line of sight:
\begin{equation}
    I(\nu)=\int\epsilon(\nu)ds.
\end{equation}

To be conclusive, the major ingredients of the model include an SDP propagation scheme, a nearby source component which mainly accounts for the spectral bumps of primary CRs, and the secondary production around acceleration sources which is to explain the B/C hardening and diffuse $\gamma$-ray excesses. Note that the spectral breaks of primary and secondary particles are likely a coincidence. With higher energy measurements of the primary spectra by DAMPE \citep{2019SciA....5.3793A,2021PhRvL.126t1102A} and CALET \citep{2019PhRvL.122r1102A,2020PhRvL.125y1102A,2023PhRvL.130q1002A}, it was shown that the break rigidity is about 500 GV for protons, about 650 GV for helium, 400 - 500 GV for carbon and oxygen nuclei. The break rigidity of B/C and B/O measured by DAMPE is about 200 GV \citep{2022SciBu..67.2162D}. Therefore, the break energies for primary and secondary nuclei may indeed have different origins.

\section{Galactic diffuse emission}
\label{gde}

Starting from the model setting described in Sec. \ref{mdl}, we adjust the model parameters to match the measured spectra of CR protons, positrons, CREs, and boron-to-carbon ratio, as shown in Figure~\ref{fig:spec} in Appendix A. The diffuse emission from neV to PeV energies is then calculated, which will be discussed in detail as follows.

\begin{figure}[htb]
\centering
\includegraphics[width=0.45\textwidth]{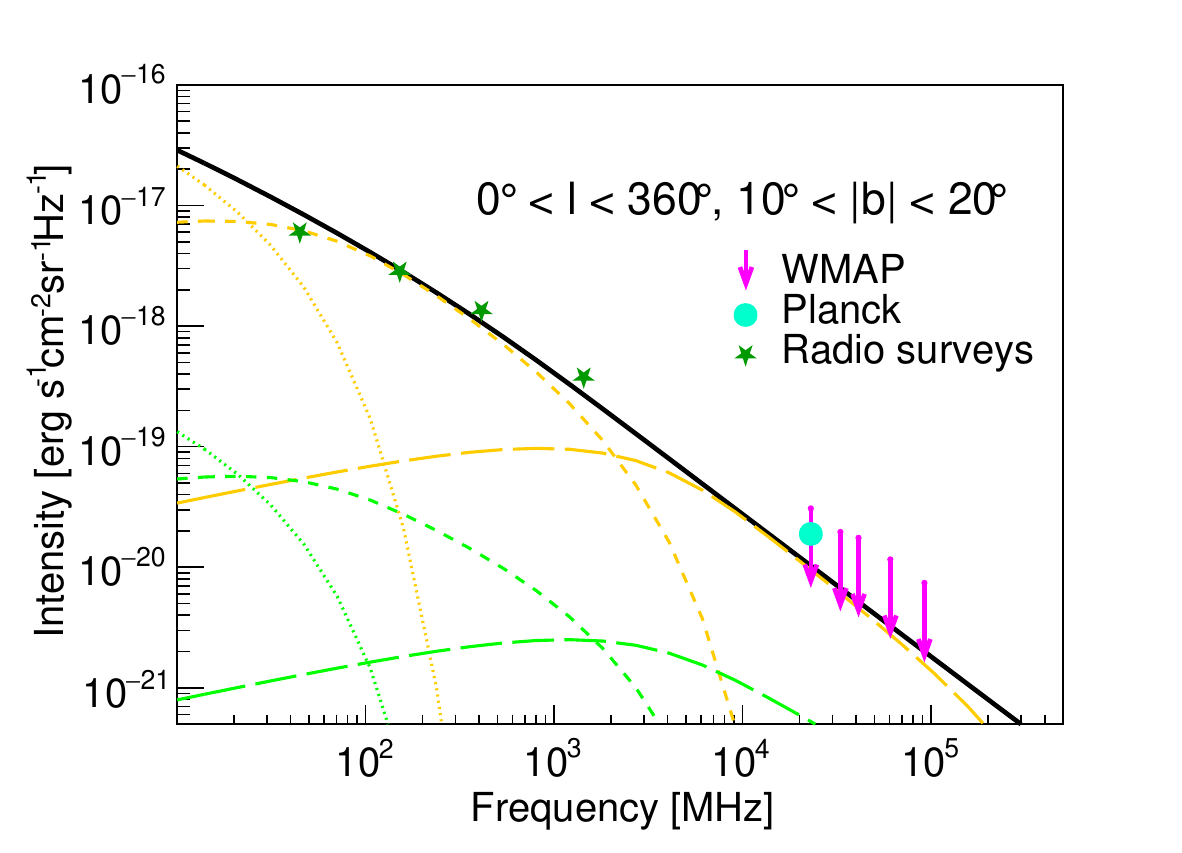}
\includegraphics[width=0.45\textwidth]{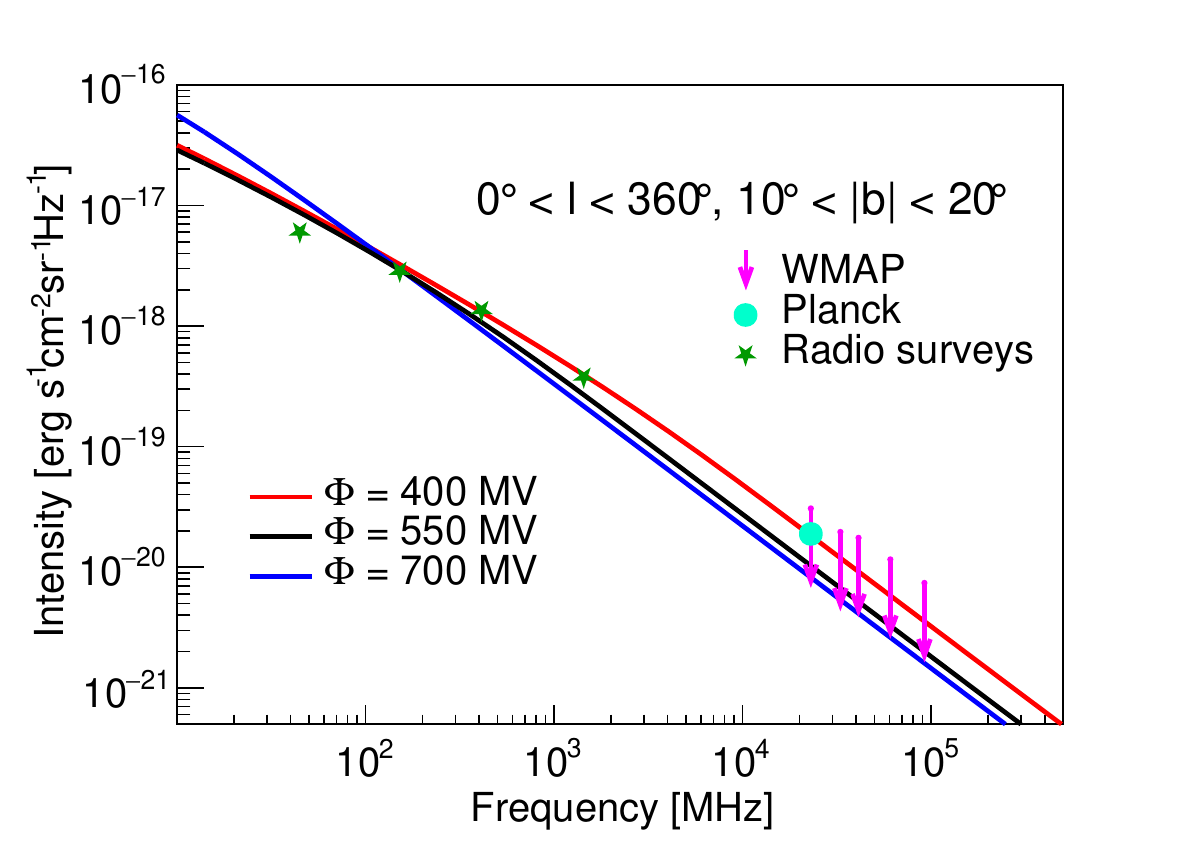}
\caption{The model predicted synchrotron emission, compared with data compiled or processed in \citet{2018MNRAS.475.2724O}, based on original measurements: radio surveys (green stars) \citep{2011A&A...525A.138G, 1970AuJPA..16....1L, 1981A&A...100..209H, 1982A&AS...47....1H, 1999ASPC..168...78R, 2001A&A...376..861R}, Planck synchrotron map (teal circle) \citep{2016A&A...594A..10P} and WMAP (magenta arrows) \citep{2013ApJS..208...20B}. In the left panel, the black solid line represents the overall intensity with modulation potential of 550 MV. The colour orange and green mark the radiation from background and fresh interactions respectively; the dotted, dashed and long-dashed lines correspond to contributions by CREs at various energy ranges: $[0.1\ \si{GeV},\ 1\ \si{GeV}]$, $[1\ \si{GeV},\ 10\ \si{GeV}]$ and $[10\ \si{GeV},\ 100\ \si{GeV}]$. In the right panel, the total synchrotron fluxes for different solar modulation potentials are shown. Note that the magnetic field strength for each case is slightly re-scaled to better fit the wide-band data.}
\label{fig:sync}
\end{figure}

\subsection{Diffuse radio emission}

The calculated synchrotron fluxes are shown in the left panel of Figure~\ref{fig:sync}, compared with the data taken from \citet{2018MNRAS.475.2724O}, for a sky region slightly above the Galactic plane ($10^\circ <|b|<20^{\circ}$). The model prediction is roughly consistent with the observations. At high frequencies the model flux is slightly lower than the Planck data. This may be solved by a slight re-scale of the magnetic field strength and lowering CRE fluxes below $\sim 1$ GeV (see discussion on the uncertainty of the solar modulation in the next paragraph). Note that the WMAP results seem to be higher than the Planck flux at similar frequency (23 GHz). As discussed in \citet{2018MNRAS.475.2724O}, there might be degeneracies among different components in such frequency bands, e.g., synchrotron, free-free, thermal dust, and anomalous microwave emissions. The WMAP synchrotron intensities may be over-estimated. We therefore use WMAP results as upper limits (shown by arrows). 

To better see the mapping relation between CREs and synchrotron emission, we show the contributions from CREs in three energy bands, $0.1-1$ GeV, $1-10$ GeV, and $10-100$ GeV, respectively. The emission below 1 GHz mainly comes from CREs with energies smaller than $1\ \si{GeV}$, in which range the fluxes are uncertain due to solar modulation. In the right panel of Figure~\ref{fig:sync}, we show the synchrotron fluxes for three different values of the modulation potentials, 550 MV as the benchmark, 400 MV, and 700 MV, respectively. For each modulation parameter, the source parameters are tuned to reproduce the measured CR spectra (see Table \ref{tab:bg_inj}). The magnetic field strength is slightly re-scaled to better fit the radio data. For $\Phi=400$ MV, the low energy CRE fluxes are lower (Figure~\ref{fig:spec}), and the resulting synchrotron spectrum is harder, which can better match the data in a wide frequency band.

\subsection{Diffuse $\gamma$-ray emission}

\begin{figure}[!htb]
\centering
\includegraphics[width=0.45\textwidth]{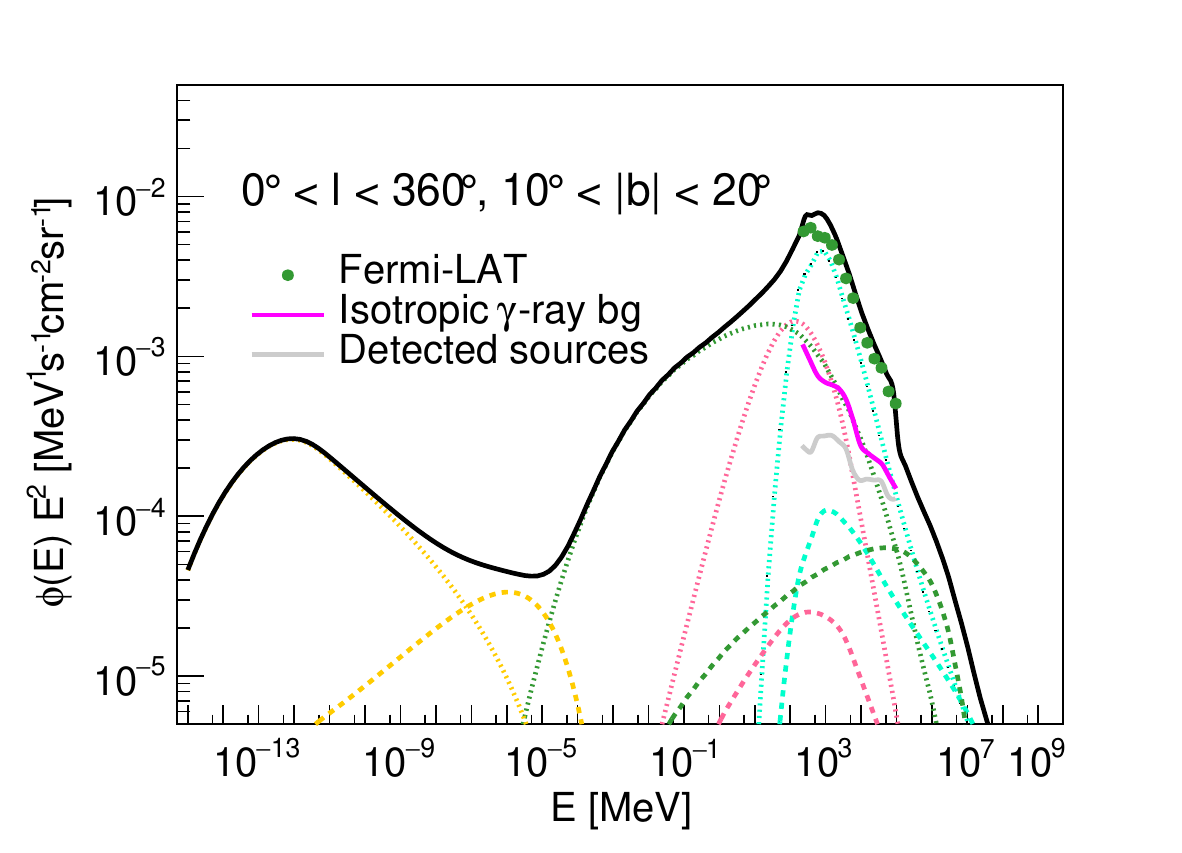}
\includegraphics[width=0.45\textwidth]{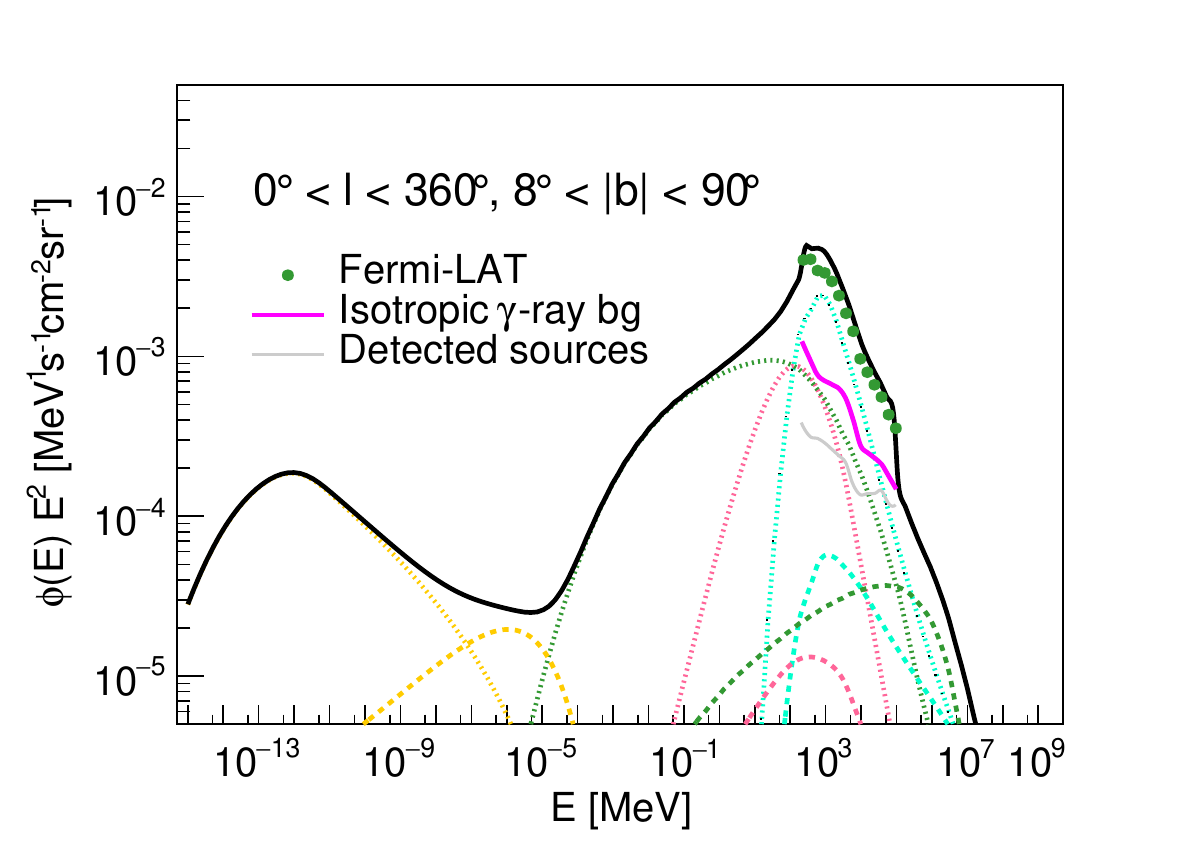}
\includegraphics[width=0.45\textwidth]{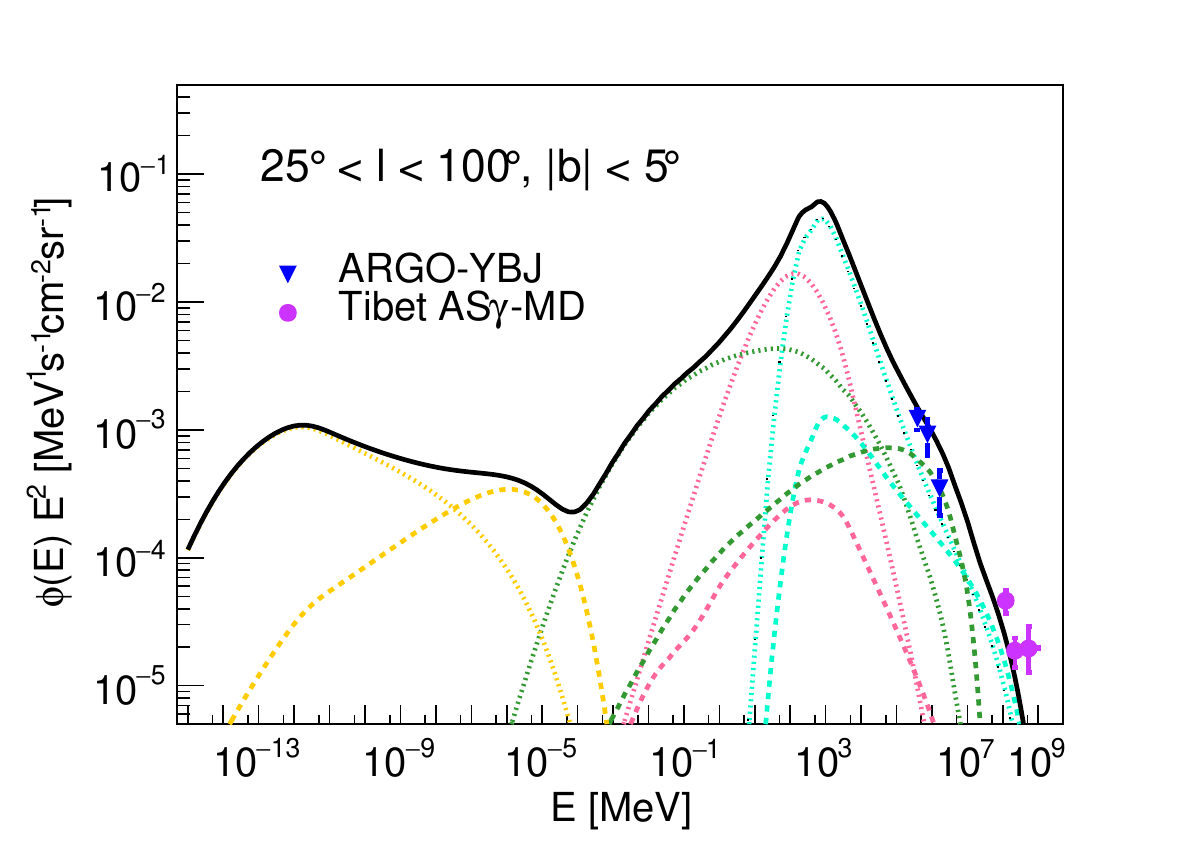}
\includegraphics[width=0.45\textwidth]{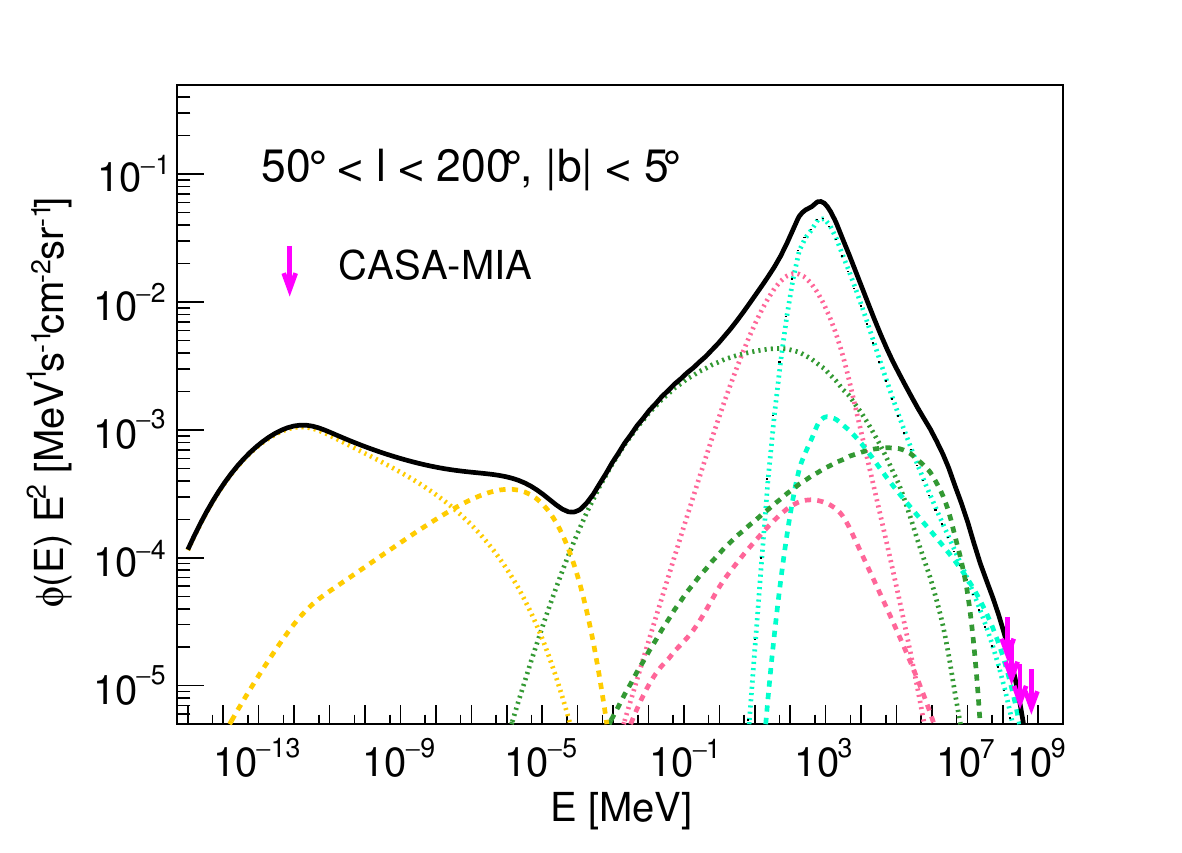}
\includegraphics[width=0.45\textwidth]{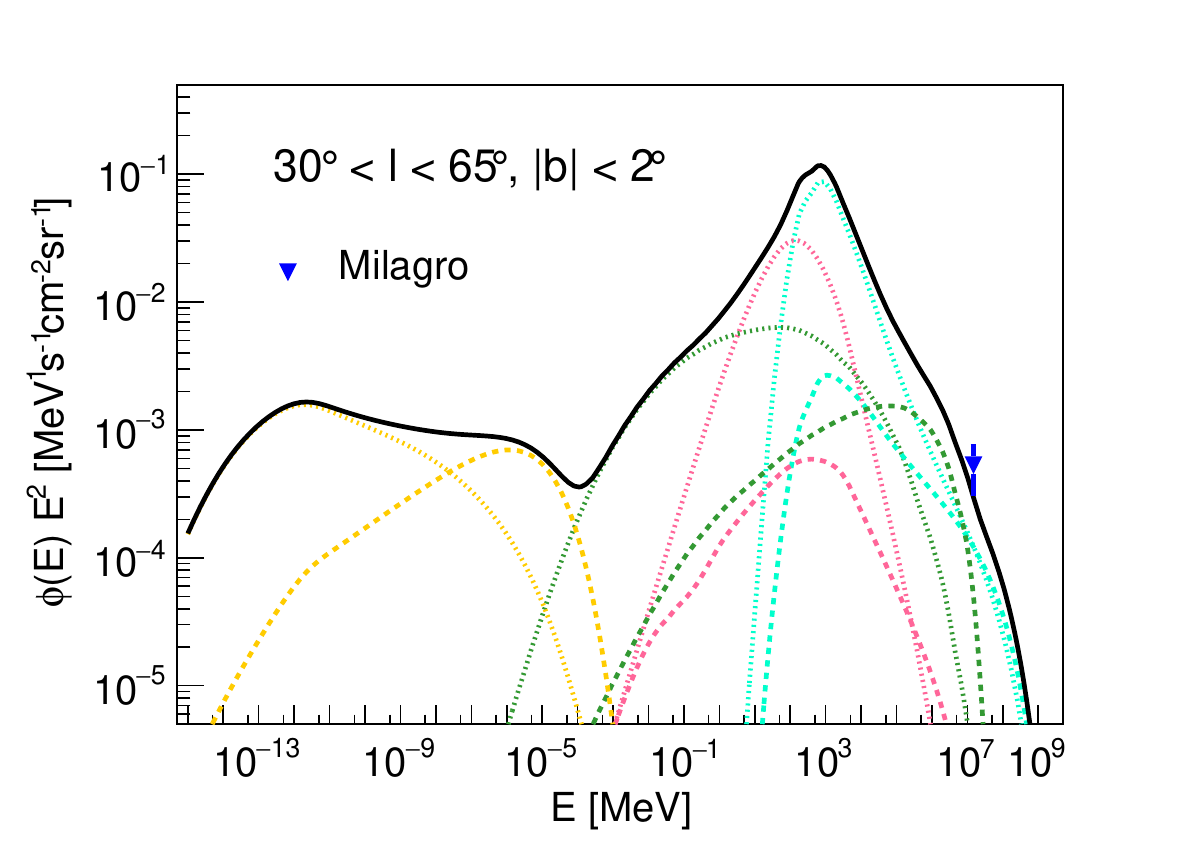}
\caption{Diffuse emissions from radio to PeV $\gamma$ rays. The black solid line is the total radiation. For the rest of the lines, the colour orange, pink, green and teal denotes components from synchrotron, Bremmstrahlung, ICS and $\pi_0$ decay; the dots and dashes tell whether they are generated from background or fresh interactions; the magenta and gray solid lines are contributions by the IGRB and resolved sources detected by Fermi-LAT. Data are from: Fermi-LAT \citep{2012ApJ...750....3A}, ARGO-YBJ \citep{2015ApJ...806...20B}, Tibet AS$\gamma$ \citep{2021PhRvL.126n1101A}, CASA-MIA \citep{1998ApJ...493..175B}, Milagro \citep{2008ApJ...688.1078A}.}
\label{fig:gde}
\end{figure}

The comparisons between model calculations and observed $\gamma$-ray data in different sky regions are shown in Figure~\ref{fig:gde}. The Fermi-LAT measurements in two latitude belts, $10\degree<|b|<20\degree$ and $8\degree<|b|<90\degree$, for the whole longitude range \citep{2012ApJ...750....3A}, the ARGO-YBJ \citep{2015ApJ...806...20B} and Tibet-AS$\gamma$ \citep{2021PhRvL.126n1101A} measurements in the inner Galactic plane region ($25\degree<l<100\degree$, $|b|<5\degree$), the Milagro measurement in a smaller region in the inner Galactic plane ($30\degree<l<65\degree$, $|b|<2\degree$) \citep{2008ApJ...688.1078A}, and the CASA-MIA upper limits a region covering mostly the outer plane ($50\degree<l<200\degree$, $|b|<5\degree$) \citep{1998ApJ...493..175B}, are employed for comparison. It is shown that in general the model can reproduce the observations relatively well. Only for $\sim$GeV energies the model fluxes are slightly higher. We keep in mind that the uncertainty of the solar modulation may take effect in such an energy range. 

\begin{figure}[htb]
\centering
\includegraphics[width=0.45\textwidth]{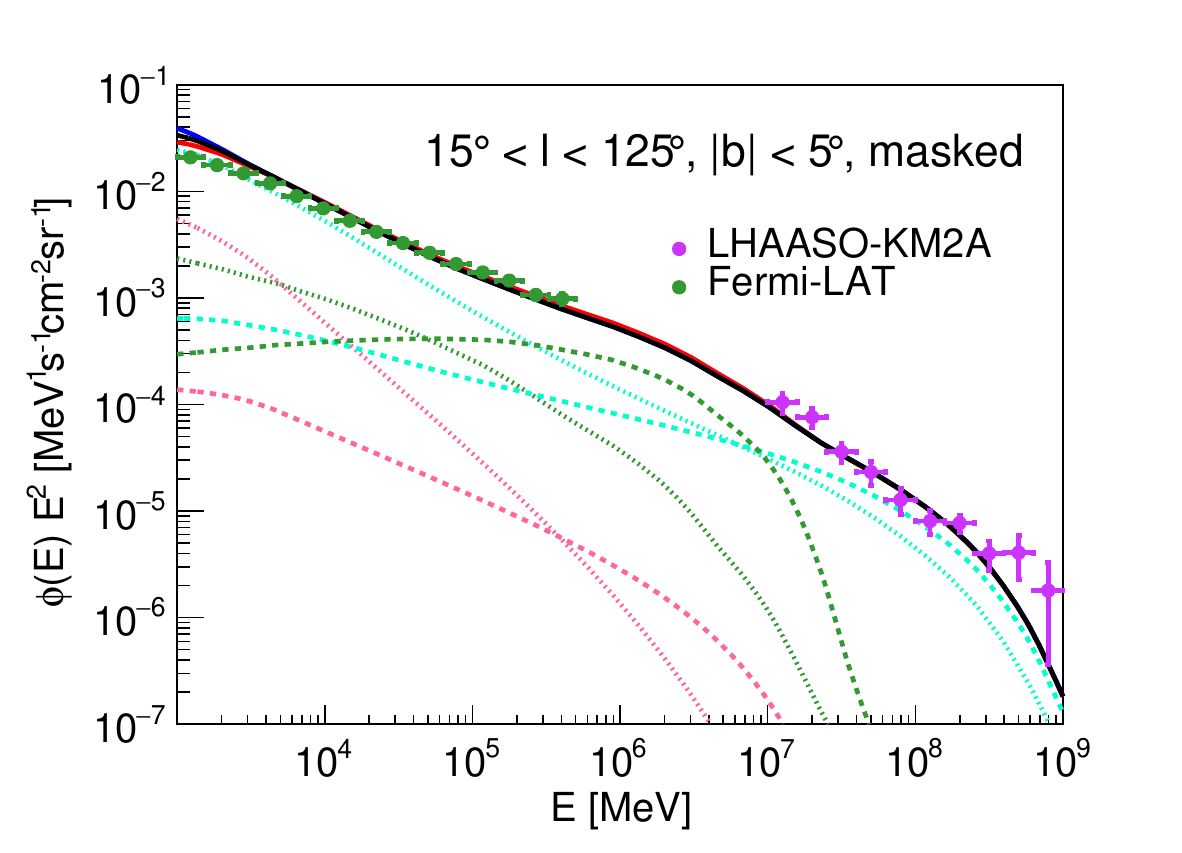}
\includegraphics[width=0.45\textwidth]{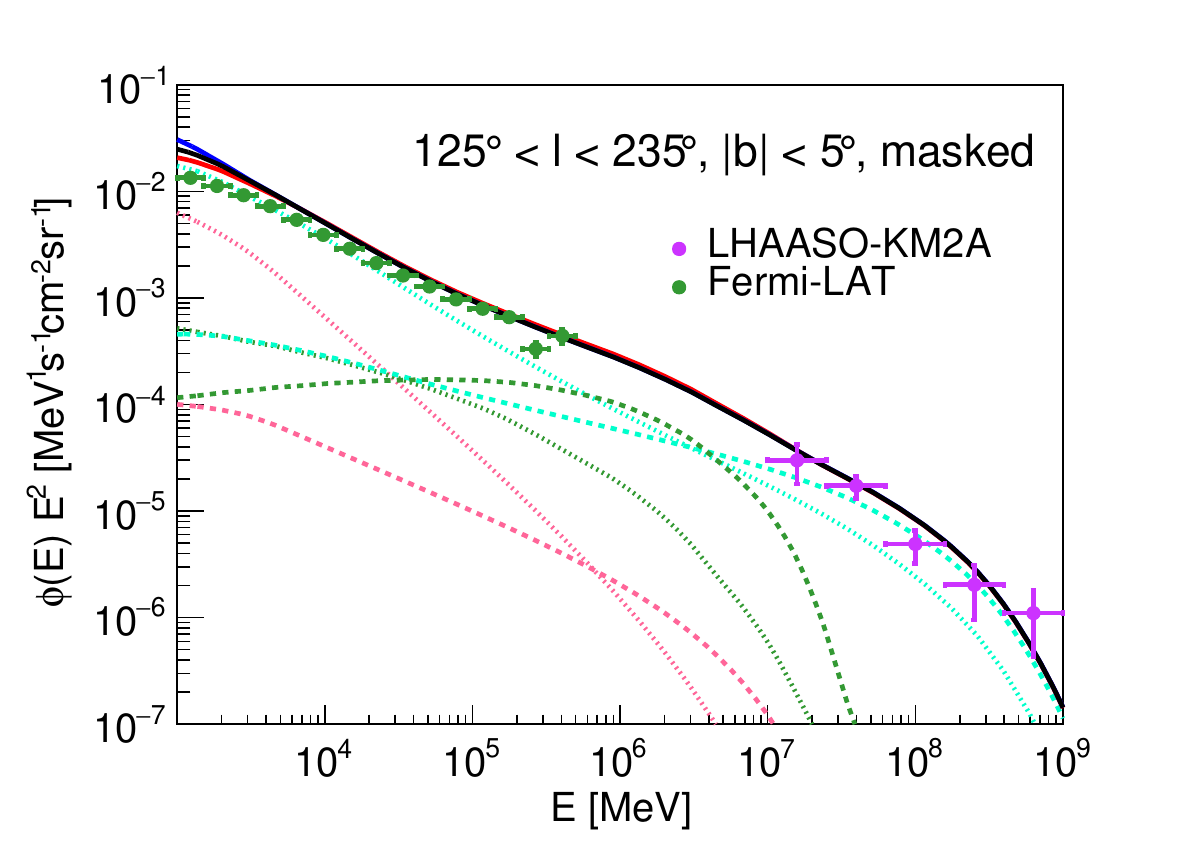}
\caption{Diffuse $\gamma$-ray emission of the inner region (left) and the outer region (right) of the Galaxy. Areas containing resolved sources detected by LHAASO-KM2A is masked, the masked region is referred to FIG. 1 in \citet{2023PhRvL.131o1001C}. The red, black and blue solid lines represent results predicted by models with modulation potentials of 400 MV, 550 MV and 700 MV. }
\label{fig:gamma}
\end{figure}

The comparison of the predicted diffuse $\gamma$-ray emission towards the Galactic plane ($|b|<5\degree$) of the inner region ($15\degree<l<125\degree$) and the outer region ($125\degree<l<235\degree$) of the Galaxy to measurements done by Fermi-LAT \citep{2023ApJ...957...43Z} and LHAASO-KM2A \citep{2023PhRvL.131o1001C} is presented in Figure~\ref{fig:gamma}.

The agreement is satisfactory to an extent: at intermediate latitudes, the contributions from GCRs and extragalactic backgrounds are comparable; at the Galactic disk, the predominant emission is from the Galactic background CR nuclei, with a small portion of electronic origins. The good consistency between the predictions and the observations of $\gamma$ rays through the sky further demonstrates the reliability of our model. In addition, the Galactic background CRs and the fresh ones make major contributions below and above tens of $\si{GeV}$s respectively. This feature has already been reflected in the hardening of the CR spectra and secondary-to-primary ratio spectra as shown in Figure~\ref{fig:spec} in Appendix A. It suffices to say that our model has been tested in the $\gamma$-ray band from tens of $\si{MeV}$s to hundreds of $\si{TeV}$s.

\subsection{Spatial distributions}

\begin{figure}
\centering
\includegraphics[width=0.3\textwidth]{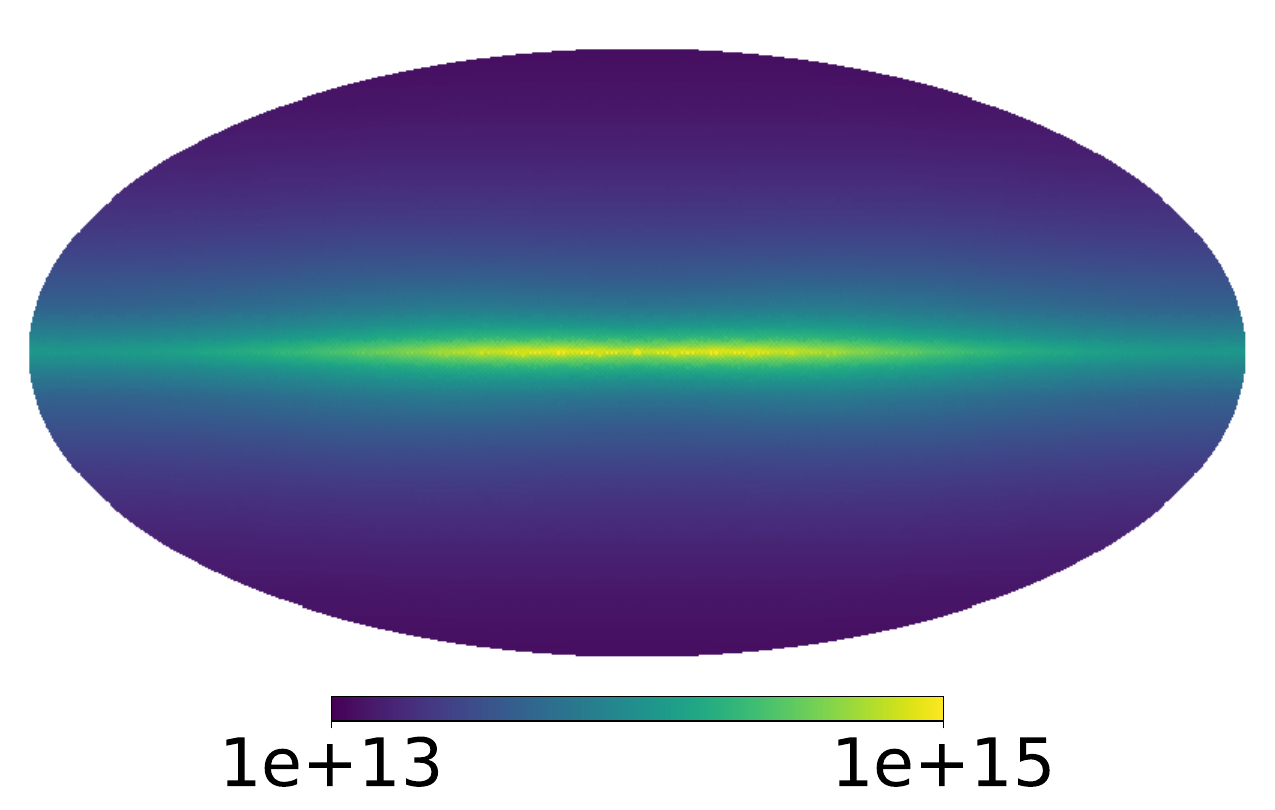}
\includegraphics[width=0.3\textwidth]{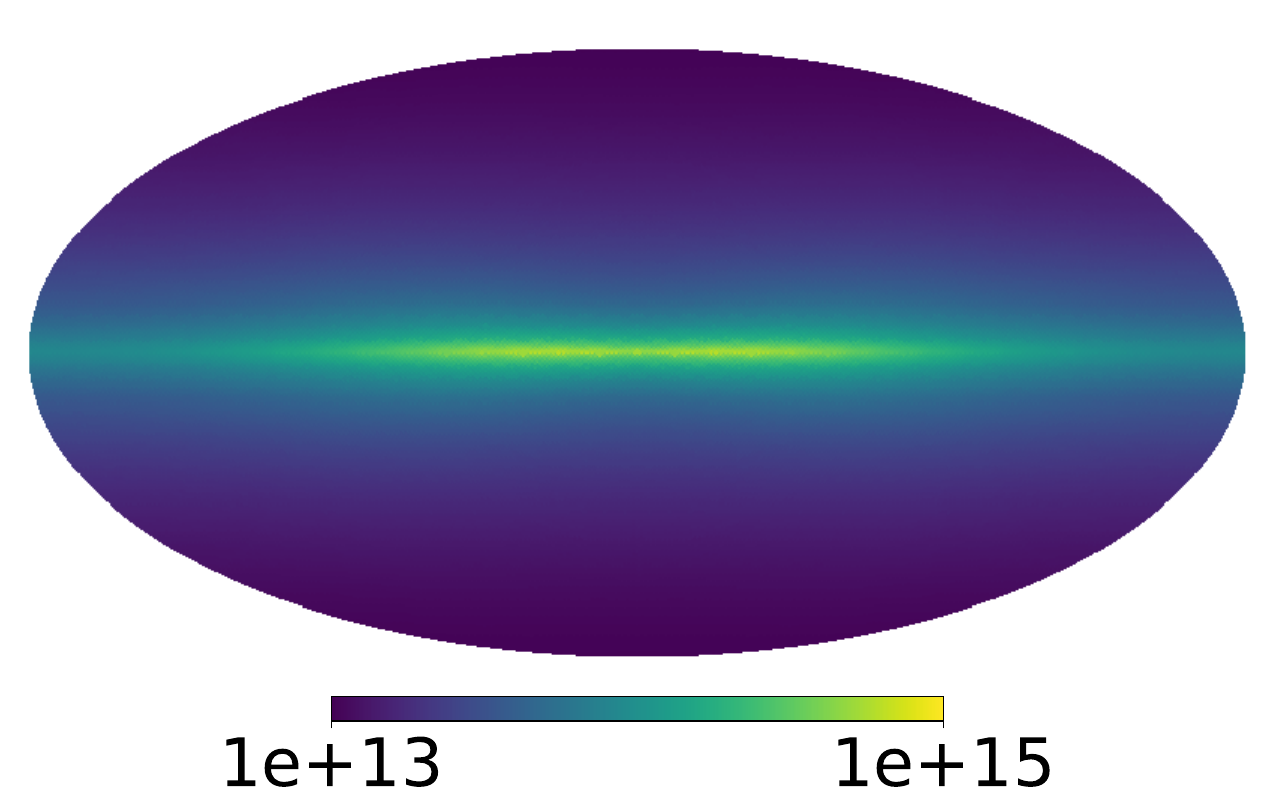}
\includegraphics[width=0.3\textwidth]{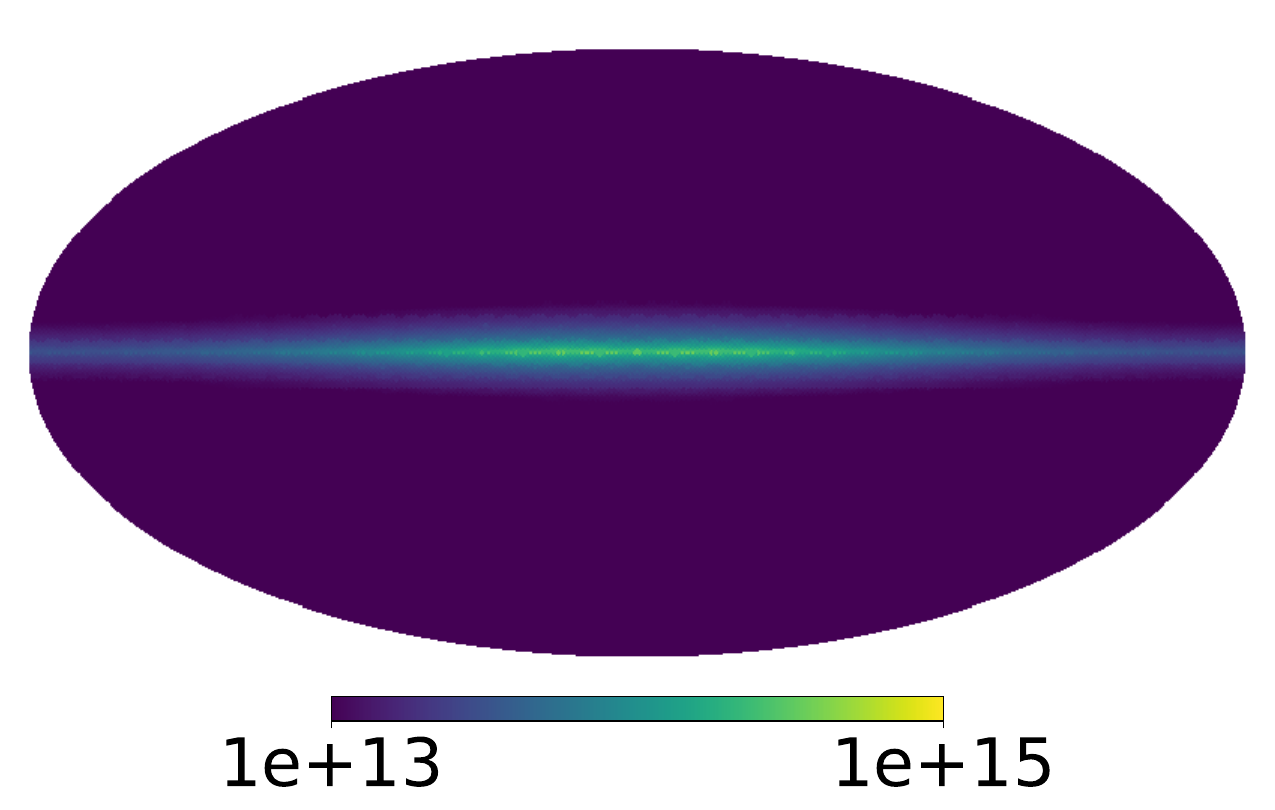}
\includegraphics[width=0.3\textwidth]{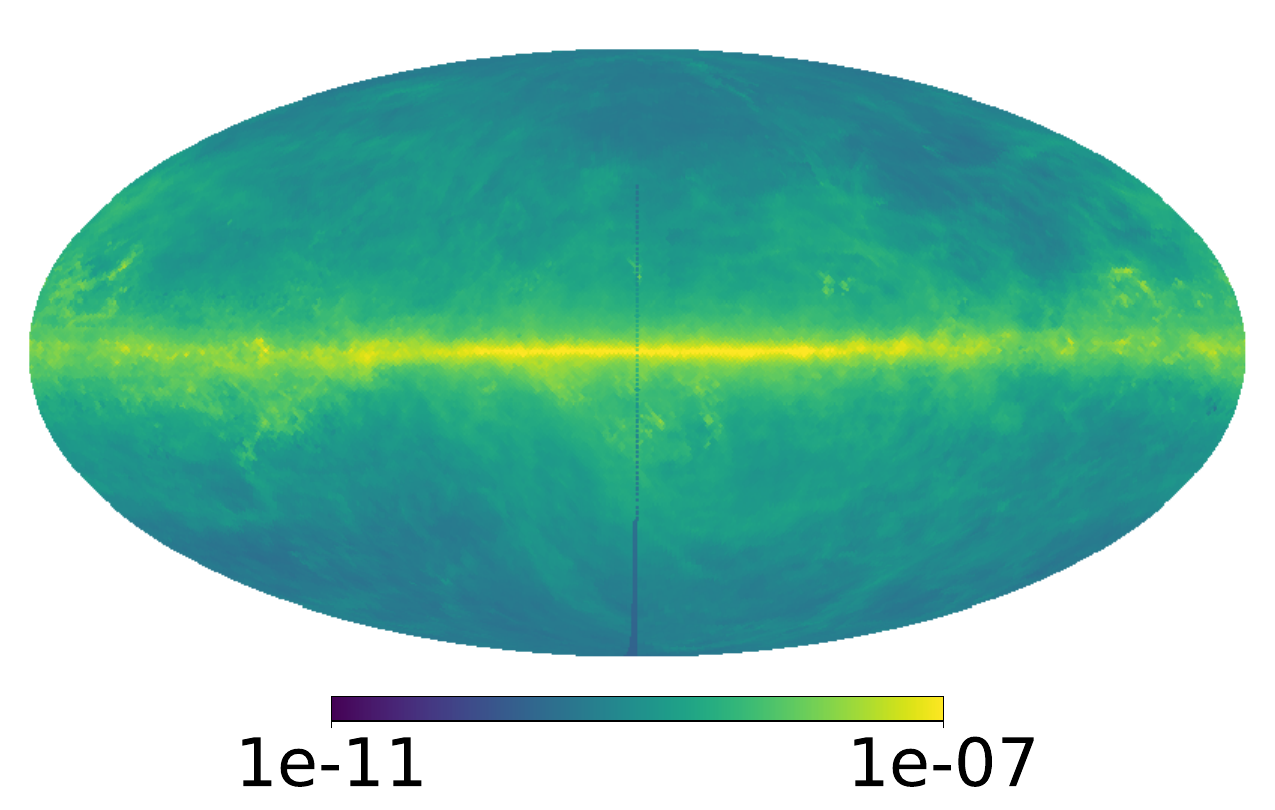}
\includegraphics[width=0.3\textwidth]{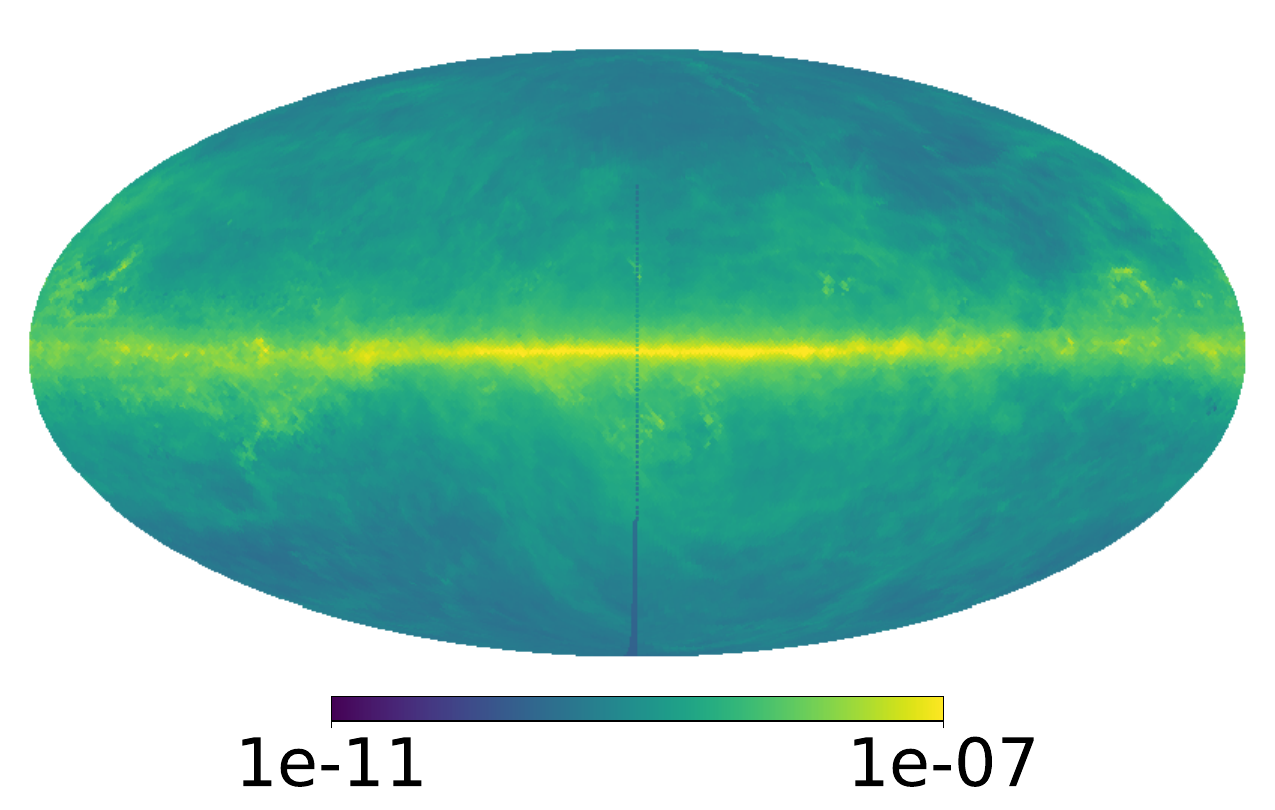}
\includegraphics[width=0.3\textwidth]{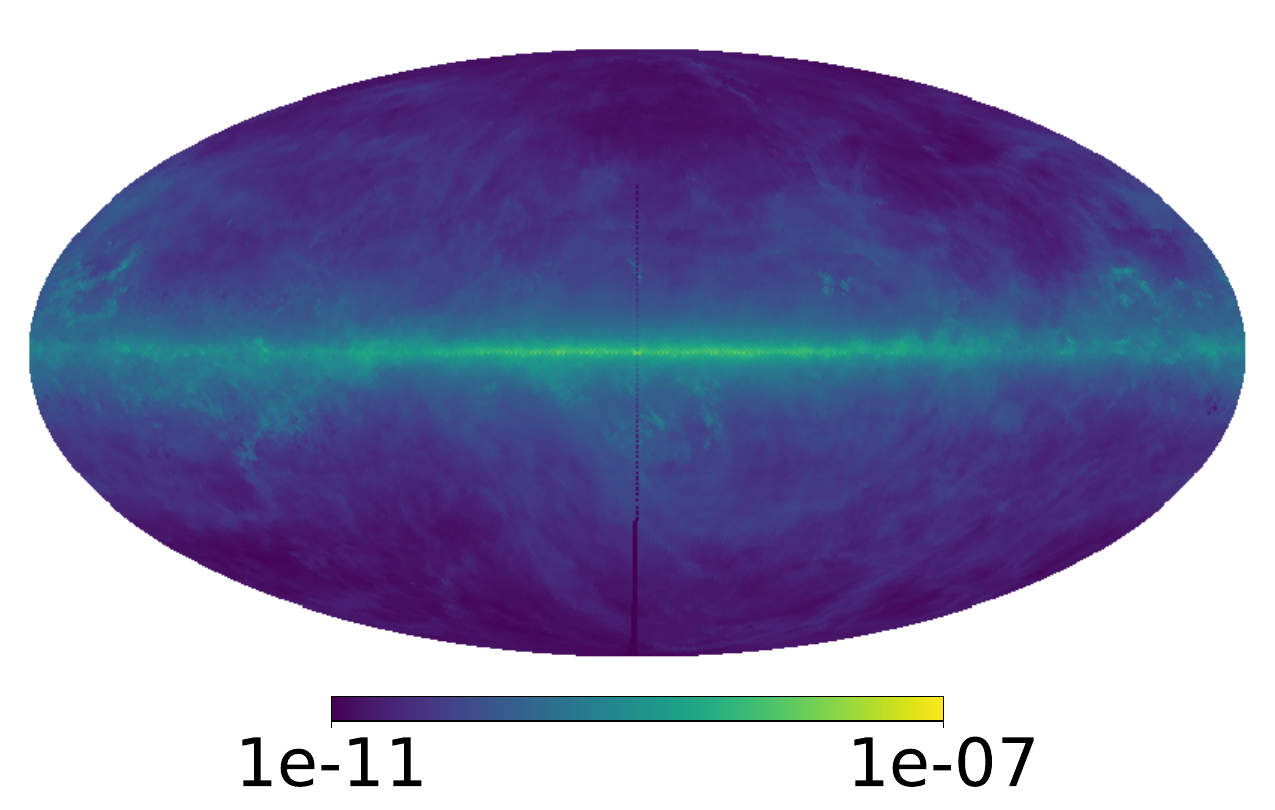}
\includegraphics[width=0.3\textwidth]{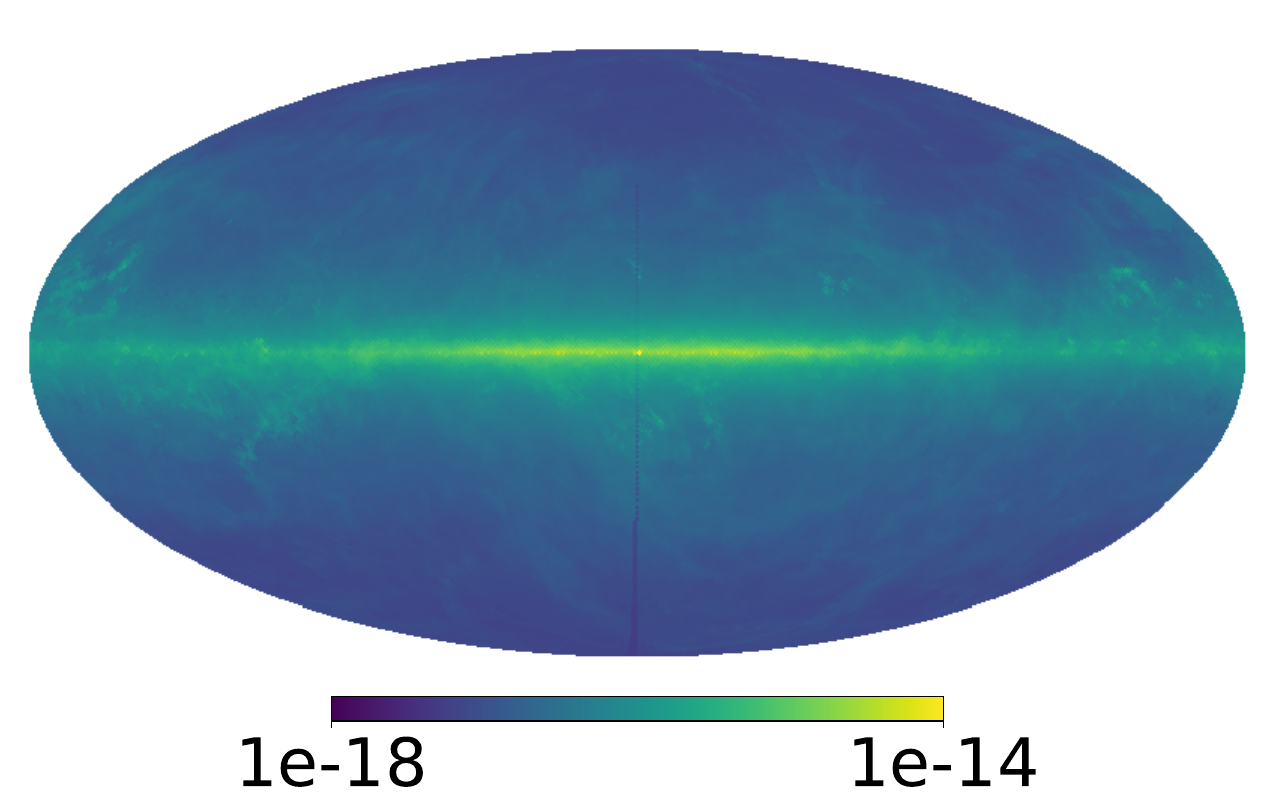}
\includegraphics[width=0.3\textwidth]{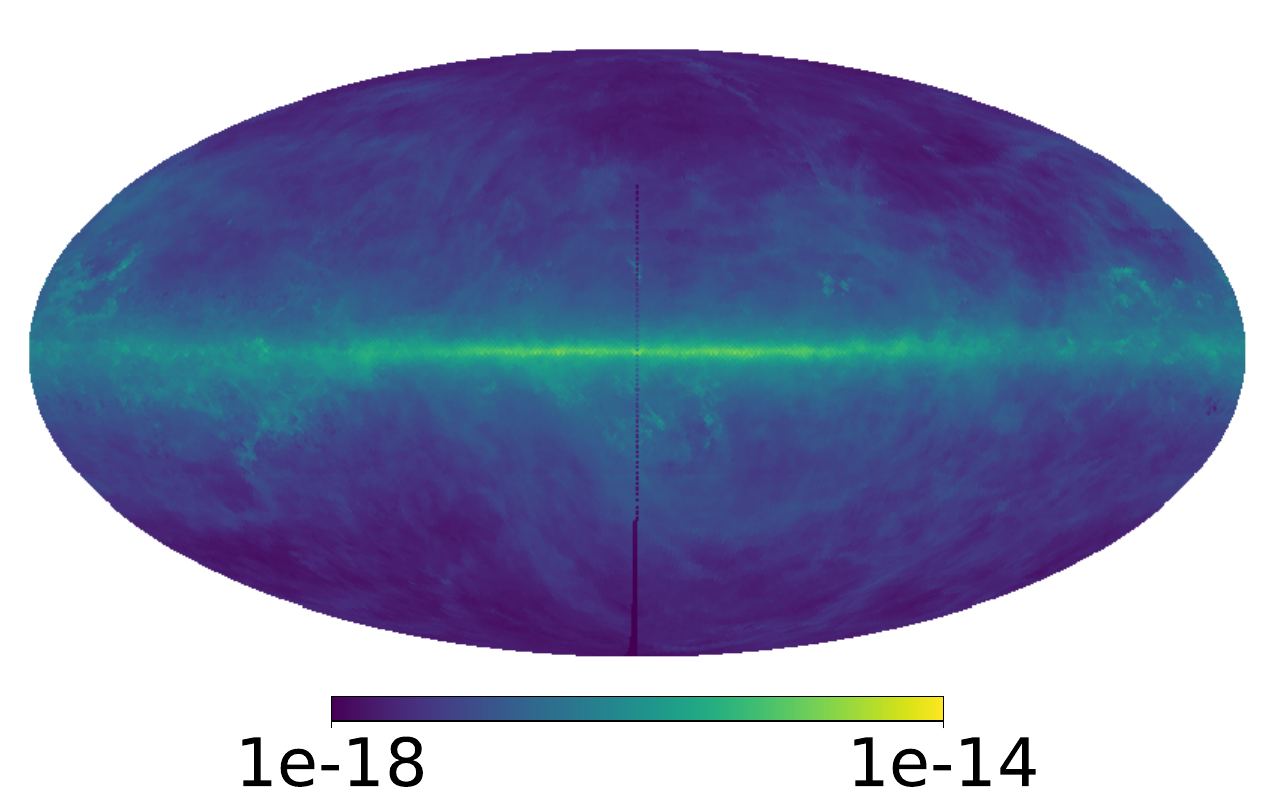}
\includegraphics[width=0.3\textwidth]{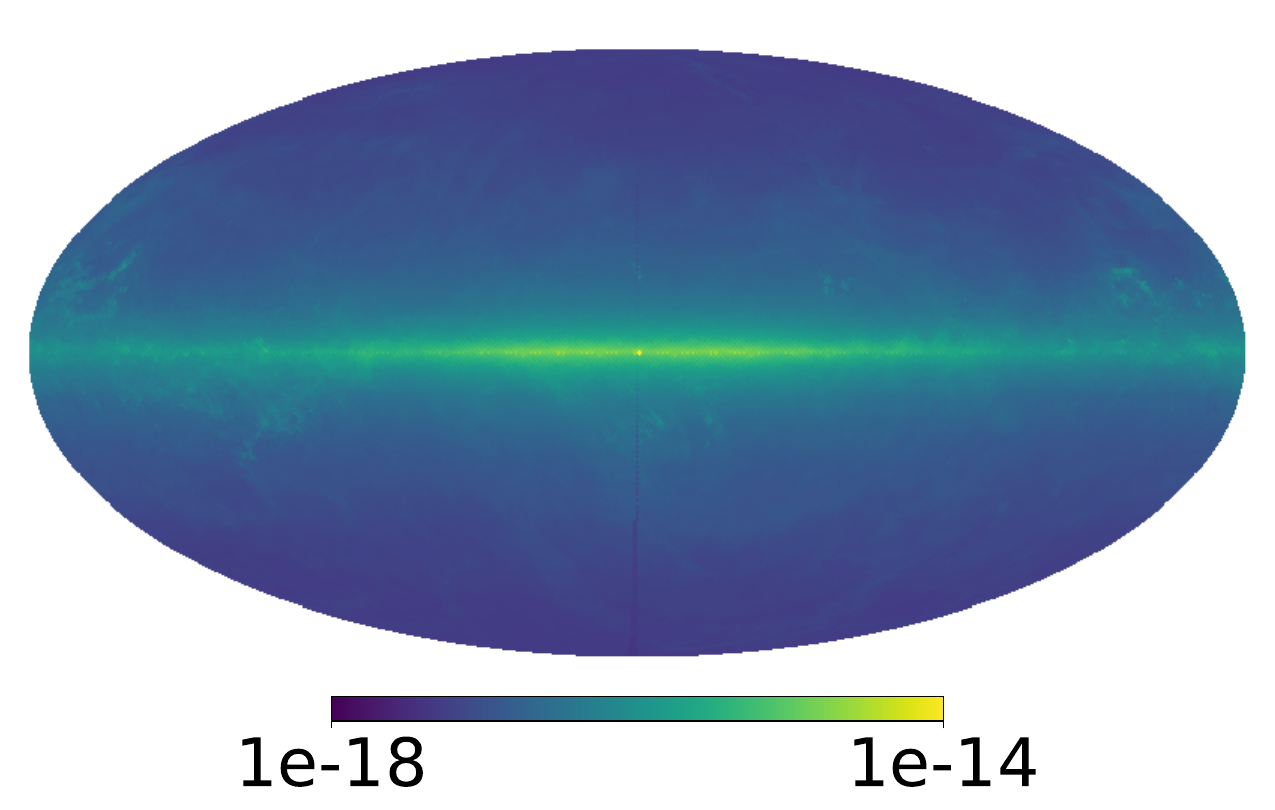}
\caption{Diffuse emission skymaps in unit of $\si{MeV^{-1}s^{-1}cm^{-3}sr^{-1}}$. From top to bottom: radiation at $1.5 \times 10^{-9}\ \si{MeV}$, $1.4 \times 10^3\ \si{MeV}$ and $1.4 \times 10^6\ \si{MeV}$; from left to right: total GDE, emission originating from background and fresh interactions. }
\label{fig:skymap}
\end{figure}

\begin{figure}[!htb]
\centering
\includegraphics[width=0.45\textwidth]{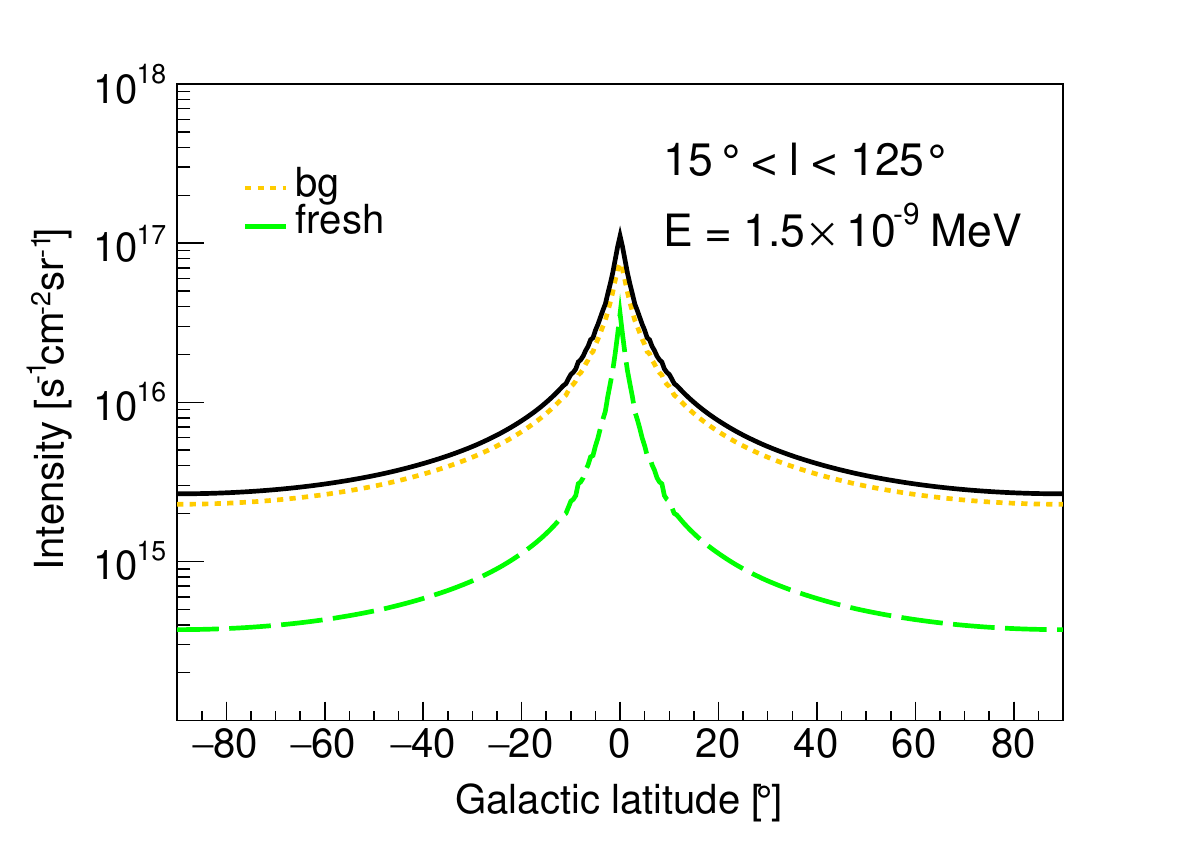}
\includegraphics[width=0.45\textwidth]{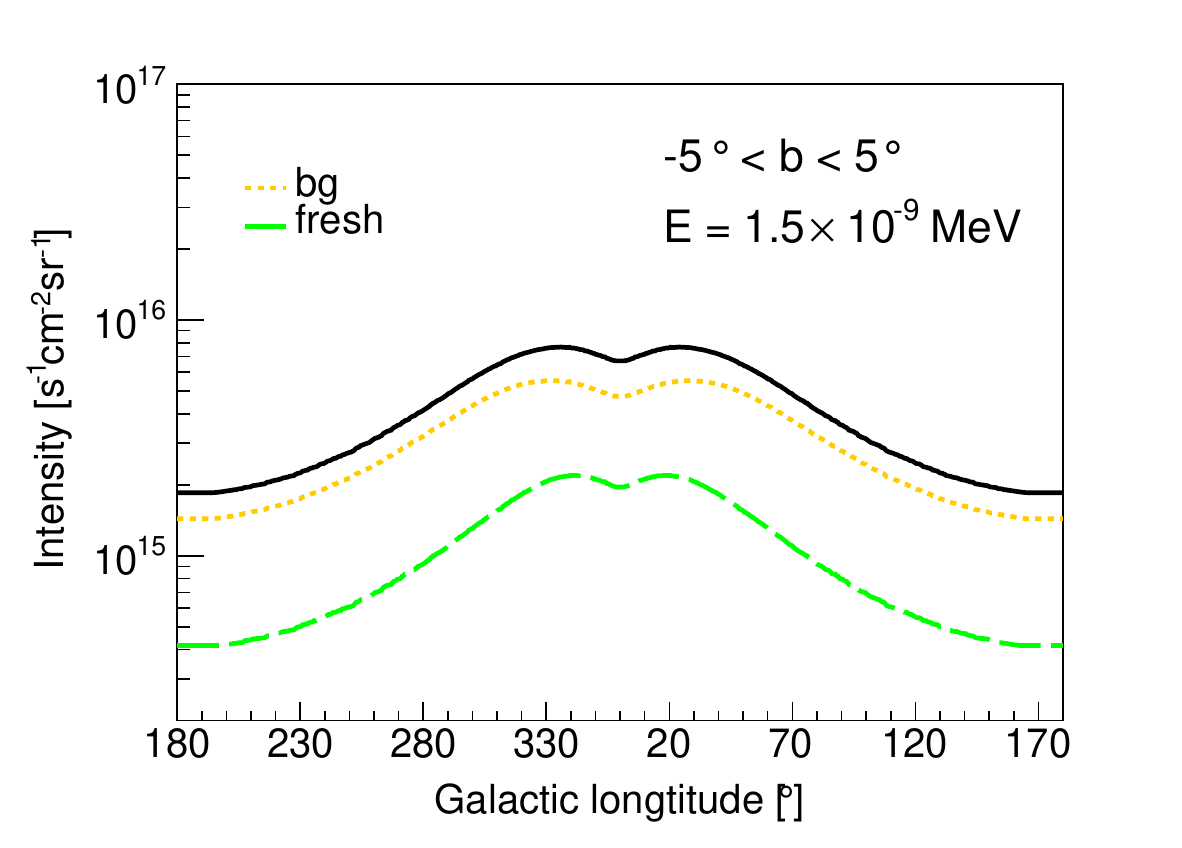}
\includegraphics[width=0.45\textwidth]{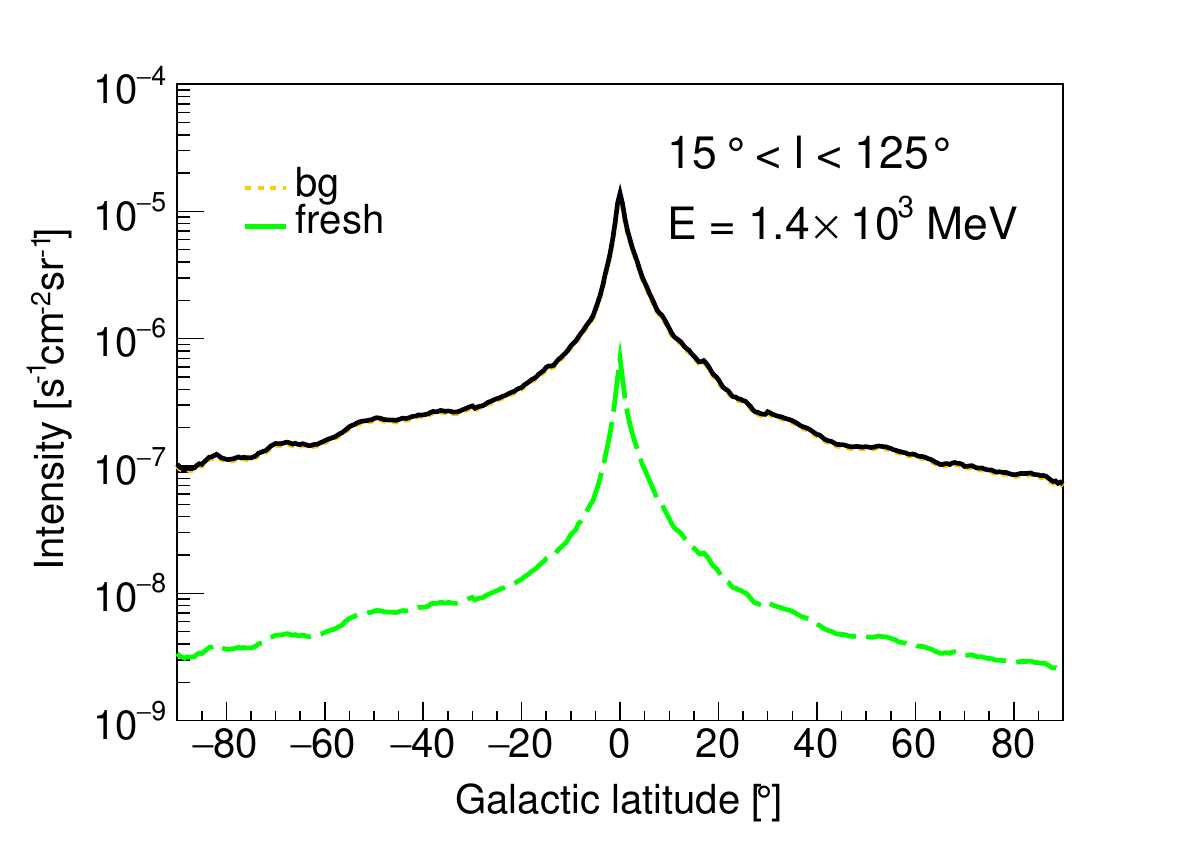}
\includegraphics[width=0.45\textwidth]{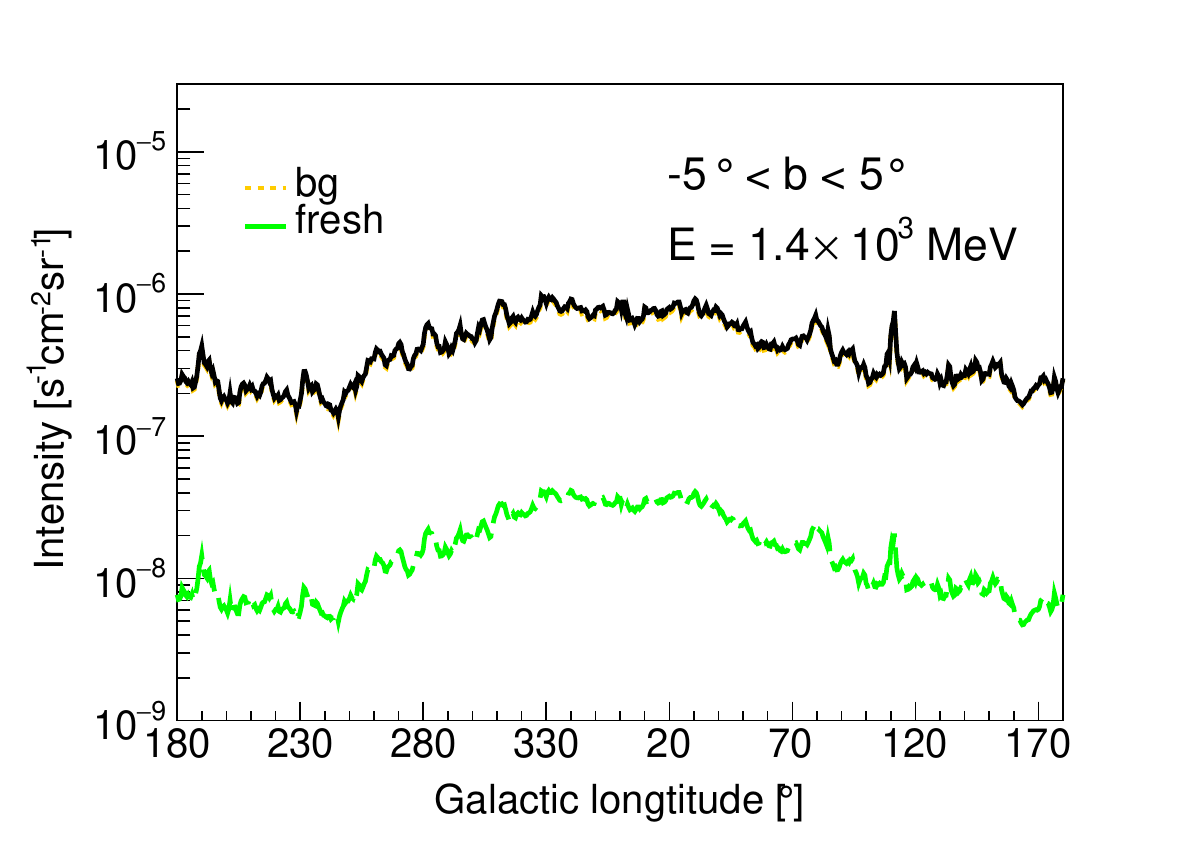}
\includegraphics[width=0.45\textwidth]{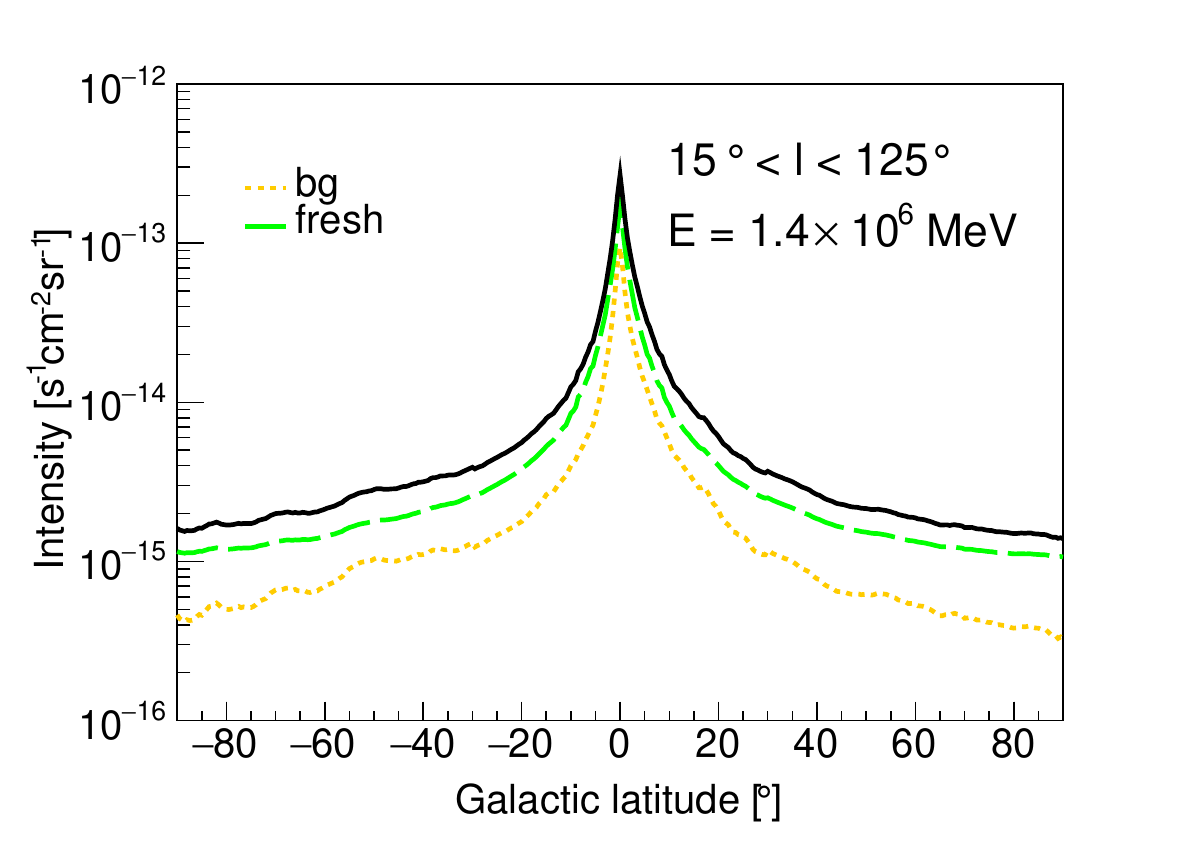}
\includegraphics[width=0.45\textwidth]{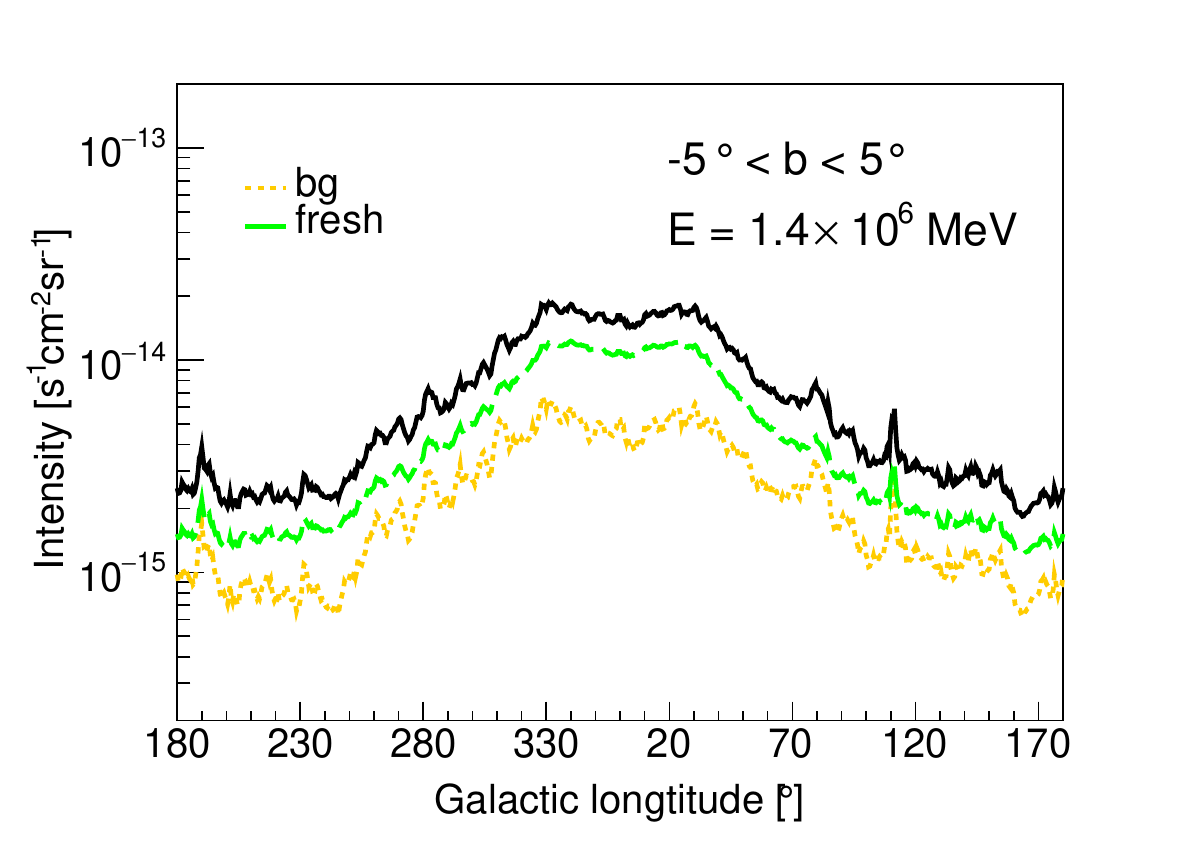}
\caption{1-d spatial distribution of the Galactic diffuse emission. From top to bottom: radiation at $1.5 \times 10^{-9}\ \si{MeV}$, $1.4 \times 10^3\ \si{MeV}$ and $1.4 \times 10^6\ \si{MeV}$; from left to right: distribution along Galactic latitudes and longtitudes. }
\label{fig:lat_long}
\end{figure}

The energy spectra from radio to PeV $\gamma$-ray emission have been calculated and compared with the measurements. There exists satisfactory consistency between the calculations and the observations in the $\gamma$-ray band over the sky. Moreover, with the aim of conducting a more comprehensive and detailed study of the CR propagation, we should delve deeper into their spatial distribution, especially those at different energy bands, to study the evolution of CR composition. Figure~\ref{fig:skymap} shows the diffuse emission skymaps for radiation at the energy of $1.5 \times 10^{-9}\ \si{MeV}$, $1.4 \times 10^3\ \si{MeV}$ and $1.4 \times 10^6\ \si{MeV}$, where two distinct features are noticeable. Firstly, the distribution appear rather smooth in the radio band and yet somewhat uneven in the range of $\gamma$ rays. Secondly, only at above $\si{TeV}$s does the fresh component dominate. To explore these features a step further, the 1-d spatial distributions along the Galactic latitudes and longitudes are explored and presented in Figure~\ref{fig:lat_long}, showing similar attributes to those exhibited already in Figure~\ref{fig:skymap}. In addition, as the injected source distribution is assumed to be identical for the CREs and CR nuclei during our calculation, the heterogeneity of the $\gamma$ rays must have originated from the distribution of the ISM. Hopefully the Fermi-LAT and the LHAASO experiments will provide cleanly subtracted diffuse skymaps in the $\gamma$-ray range and the space-borne experiments will give similar measurements in the radio and X-ray bands in the future.

\subsection{Comparison with other models}

\begin{figure}[!htb]
\centering
\includegraphics[width=0.32\textwidth]{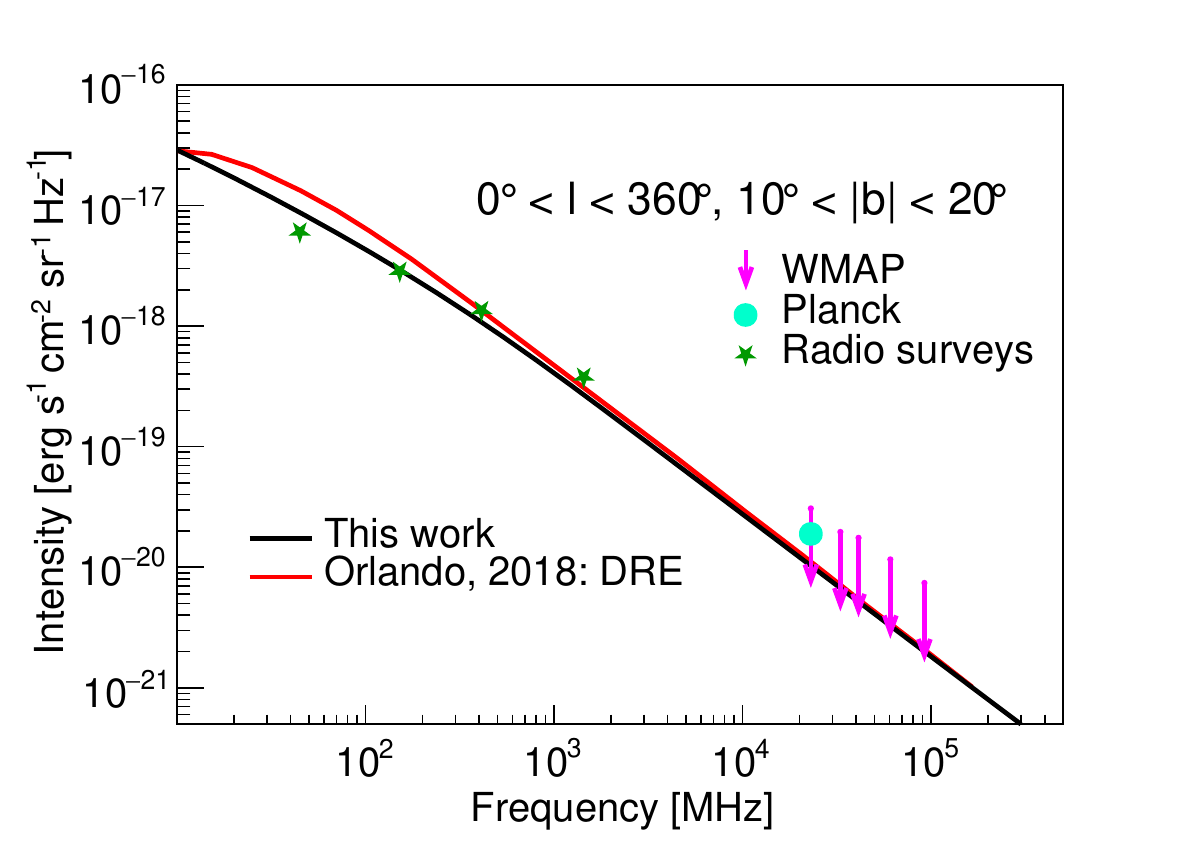}
\includegraphics[width=0.32\textwidth]{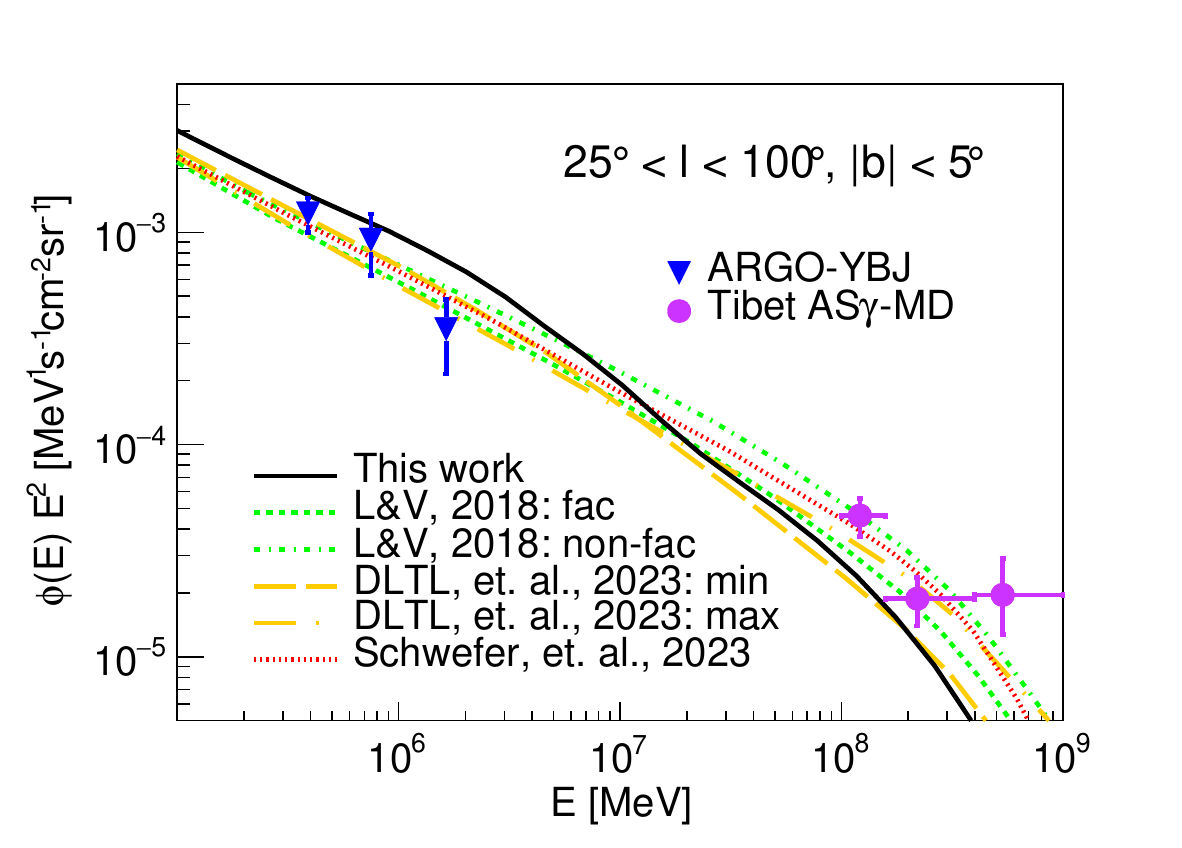}
\includegraphics[width=0.32\textwidth]{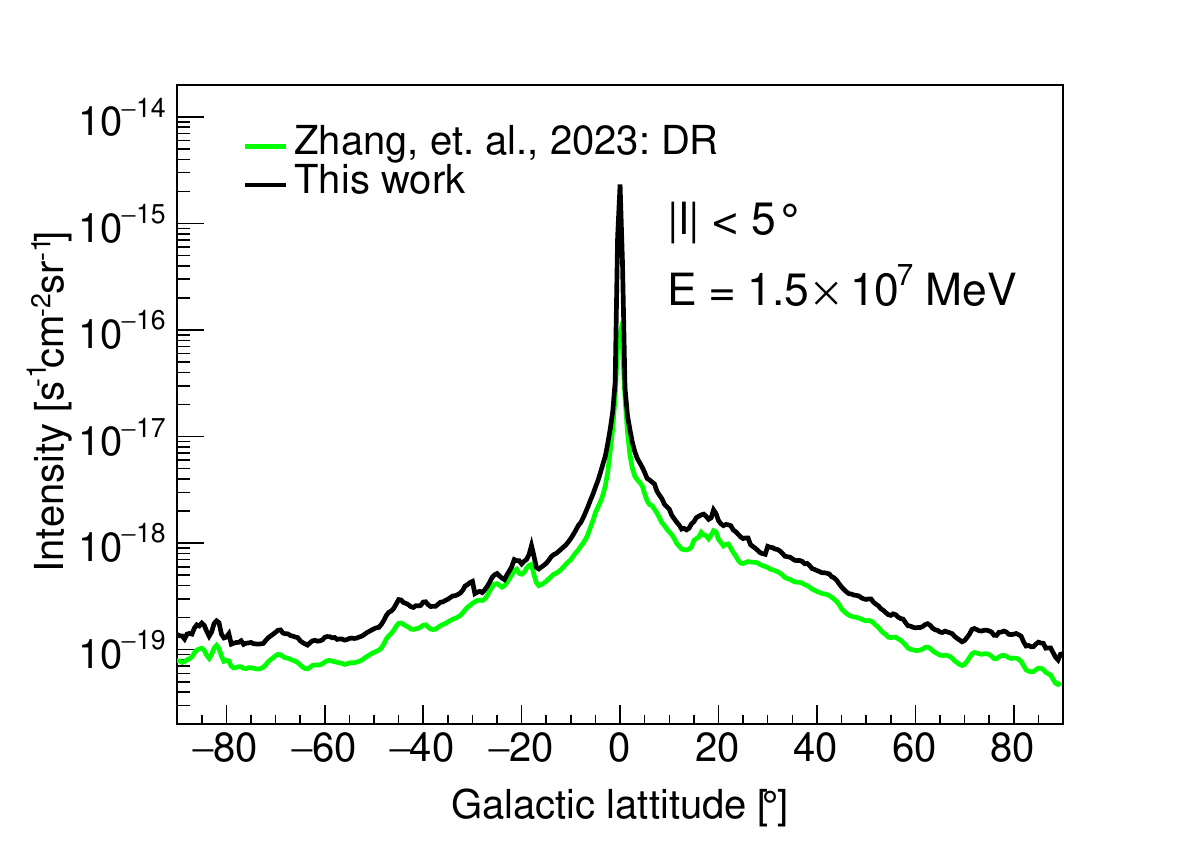}
\caption{Comparison of the results of this work to those of others \citep{2018MNRAS.475.2724O, 2018PhRvD..98d3003L, 2023A&A...672A..58D, 2023ApJ...949...16S, 2023ApJ...957...43Z}. Left: spectra of synchrotron radiation; middle: fluxes of diffuse $\gamma$-ray emission; right: Galactic latitude distribution of diffuse $\gamma$-ray emission at $1.5\times 10^7\ \si{MeV}$ energy.}
\label{fig:comparison}
\end{figure}

Similar to ours, modifications on the diffusion coefficient have been performed in other studies, e.g., a linear dependence on the distance to the Galactic center \citep{2015PhRvD..91h3012G}. Alternatively, despite minimized by our model, the influence of processes such as reacceleration and convection can also be highlighted \citep[e.g.][]{2018MNRAS.475.2724O, 2022FrPhy..1764501Q}. The unresolved $\gamma$-ray sources have also been brought up to explain the high-energy GDE \citep{2022CmPhy...5..161V,2023ApJ...949...16S}, but the correspondent radiation mechanism remain to be further understood. Some models take account of variants less recognized in this context, e.g., a spatially dependent $\mathrm{X_{CO}}$ \citep{2015PhRvD..91h3012G}, with uncertain factors such as Galactic chemical evolution. Predictions from some of these models raised above are drawn together with those from this work, as shown in Figure~\ref{fig:comparison}. At radio frequencies, we compare with the baseline DRE model of \citet{2018MNRAS.475.2724O} instead of the best-fit DRElowV model which employed some arbitrary tuning of the Alfven velocity for particular particle species. Our model gives relatively lower fluxes at low frequencies, and can better match the data. This is primarily due to that the reacceleration in our case is not as strong as the DRE model of \citet{2018MNRAS.475.2724O}. Note that the simplified magnetic field model we adopt in this work may affect the results as explored in \citep{2013MNRAS.436.2127O}. As for the diffuse $\gamma$-ray emission, previous works tend to give higher fluxes above 100 TeV, to fit the AS$\gamma$ data \citep{2021PhRvL.126n1101A}. Updated fitting to the LHAASO measurements may give slightly lower fluxes at the UHE band \citep{2023PhRvL.131o1001C}. Our model prediction differs from those works mainly in the TeV region, featuring by a bump-like structure of our model prediction, which could be tested with future measurements by LHAASO \citep{Li:2023dpg}. The one-dimensional distributions along Galactic latitudes are compared with the results of a homogeneous diffusion model of \citet{2023ApJ...957...43Z}. The results show that the SDP model in this work gives a faster decrease from the disk to the pole regions, due to a slower diffusion in the Galactic disk.

\section{Conclusions}
\label{cnc}

In this work, we have studied the multi-wavelength diffuse emission from Galactic CR interactions in a framework featuring an SDP scenario, a local CR source, together with confinement and interactions around CR accelerators, which has been proposed recently to explain a variety of phenomena in observations of CRs and $\gamma$ rays. The model predicted diffuse fluxes are found to be roughly consistent with measurements in radio and GeV-PeV $\gamma$ rays in selected sky regions; the spatial distributions of the emission are presented. These predictions largely match the observations with some disagreement either resulting from problematic data or can be improved as discussed as follows, validating the setup of our model overall.

In spite of the above conclusions we have reached, there are several points of the current study that can possibly be improved in prospect. Phenomenology wise, more comprehensive comparison between model predictions and data can be made: including SEDs covering more wavelengths (e.g., X-ray) and the spatial distributions, as new observation results become accessible. The most recent gas surveys, among other updates of astrophysical knowledge can be adopted. Additional uncertainties from e.g., cross sections, the solar modulation, and the IMF could also be taken into account. We expect with a more detailed and canonical analysis we will be able to alleviate the tensions in the parameterizations and help us obtain better understandings towards correspondent astrophysical ingredients.

\section*{Acknowledgements}
This work is supported by the National Natural Science Foundation of China (Nos. 12220101003, 
12275279, U2031110) and the Project for Young Scientists in Basic Research of Chinese 
Academy of Sciences (No. YSBR-061).


\bibliographystyle{apj}
\bibliography{main}


\appendix

\setcounter{figure}{0}
\renewcommand\thefigure{A\arabic{figure}}
\setcounter{table}{0}
\renewcommand\thetable{A\arabic{table}}

\section{CR spectra}
Here we present the flux of positrons, CREs, protons, and the boron-to-carbon ratio, according to which the model parameters are determined.

\begin{figure*}[!htb]
\centering
\includegraphics[width=0.45\textwidth]{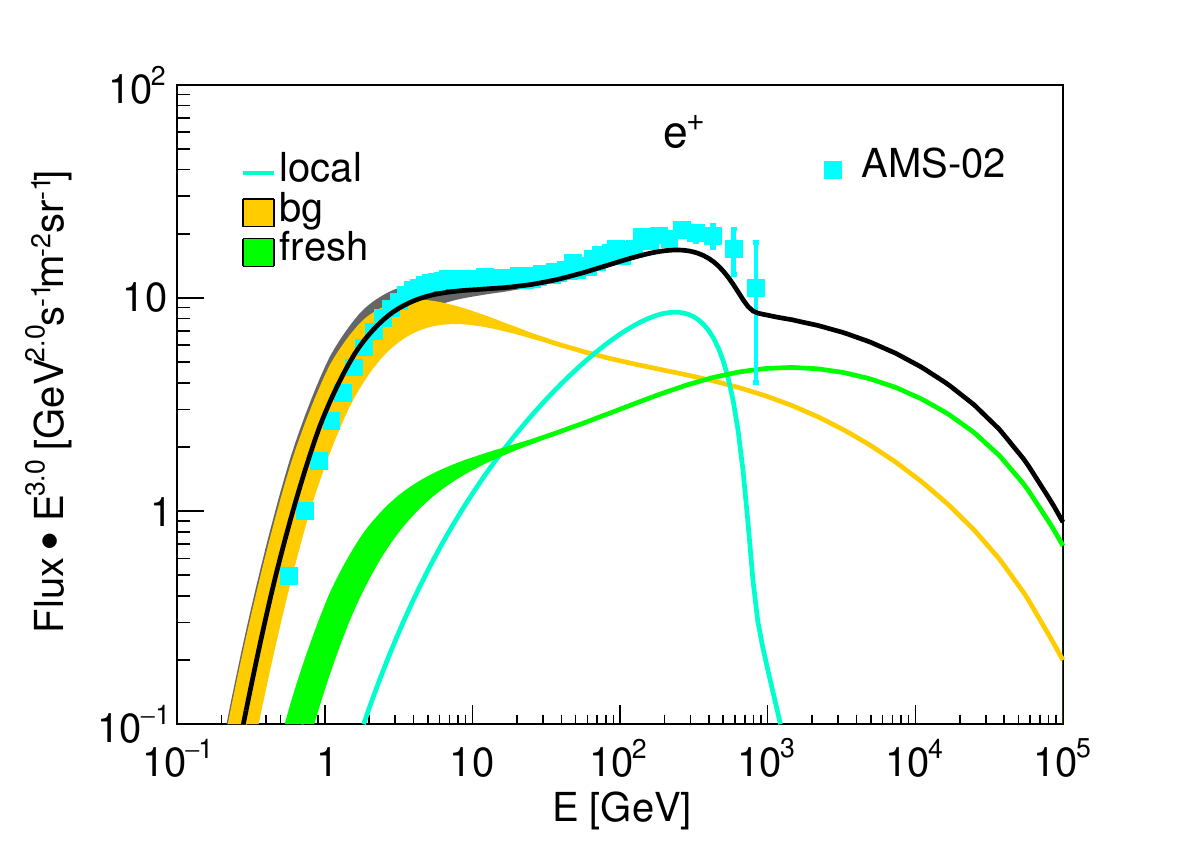}
\includegraphics[width=0.45\textwidth]{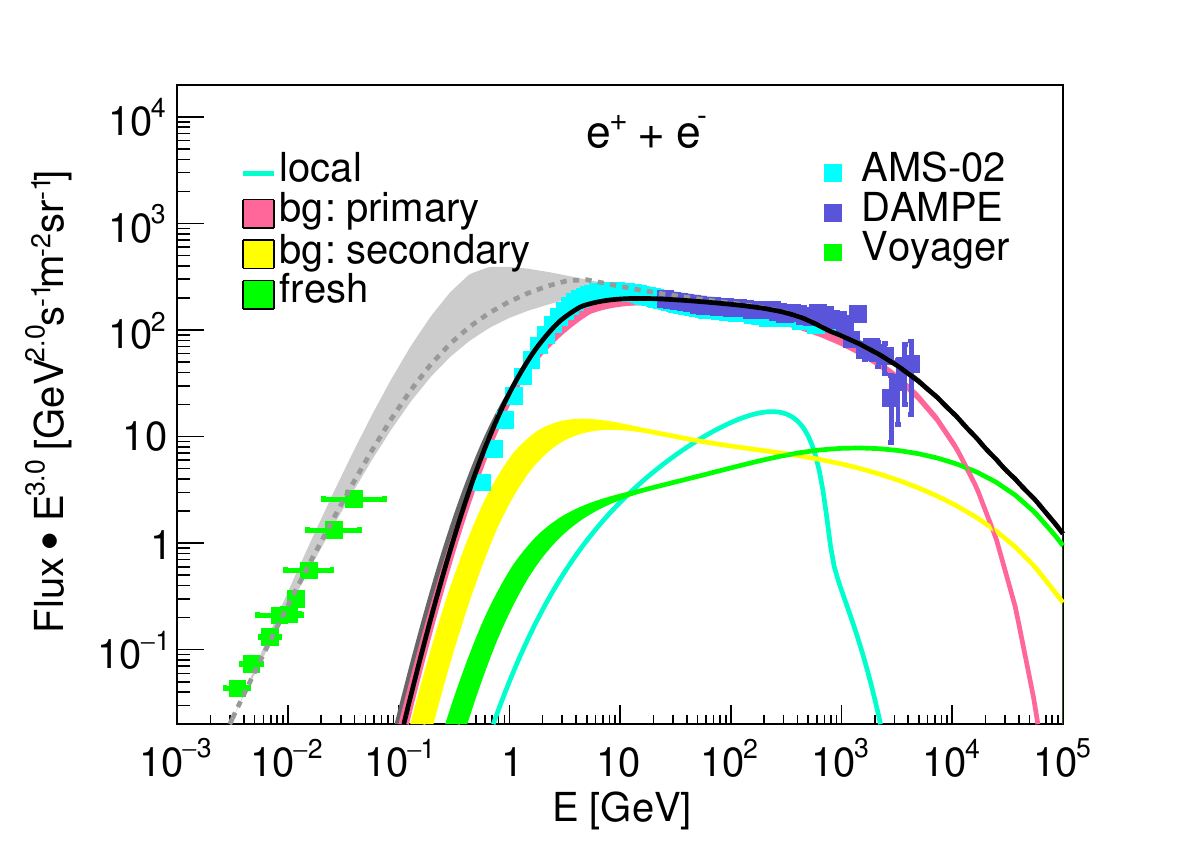}
\includegraphics[width=0.45\textwidth]{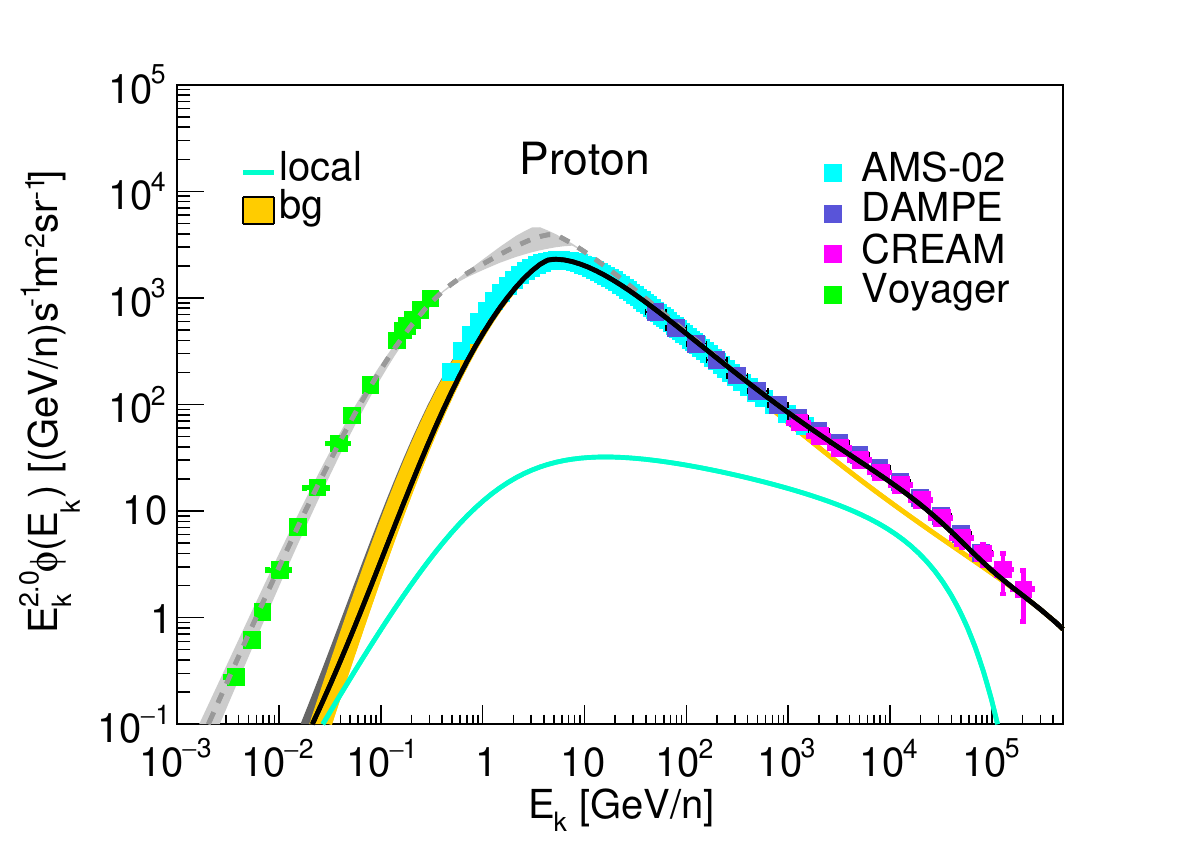}
\includegraphics[width=0.45\textwidth]{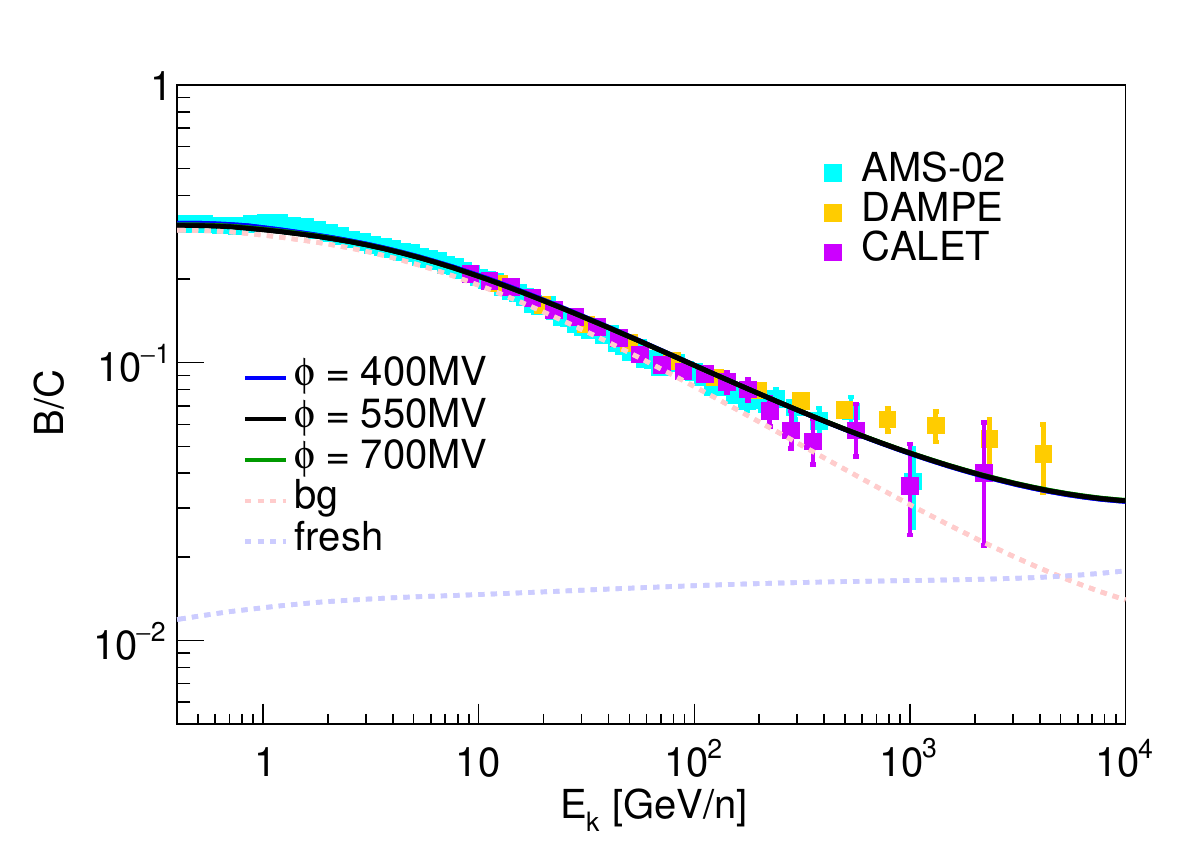}
\caption{Spectra of CR positrons (top left), electrons plus positrons (top right), protons (bottom left), and boron-to-carbon ratio (bottom right). The solid and dashed lines separate the modulated results from the unmodulated ones. For each component and the total flux, the shaded area represents the uncertainty from solar modulation parameter. Data are from experiments of Voyager \citep{2016ApJ...831...18C}, AMS-02 \citep{2019PhRvL.122d1102A,2019PhRvL.122j1101A,2015PhRvL.114q1103A,2018PhRvL.120b1101A}, CREAM \citep{2017ApJ...839....5Y} and DAMPE \citep{2017Natur.552...63D,2022SciBu..67.2162D,2019SciA....5.3793A}, and CALET \citep{2022PhRvL.129y1103A}.}
\label{fig:spec}
\end{figure*}


\label{lastpage}
\end{document}